\documentclass[twocolumn]{aastex6}

\pdfoutput=1
\usepackage{natbib}
\usepackage{amsmath}
\usepackage{multirow}
\usepackage{graphicx}
\usepackage{color}
\usepackage{threeparttable}
\usepackage{enumitem}
\usepackage{lipsum}

\newcommand\blfootnote[1]{%
  \begingroup
  \renewcommand\thefootnote{}\footnote{#1}%
  \addtocounter{footnote}{-1}%
  \endgroup
}

\def\igr{IGR J17091--3624}
\def\grs{GRS~1915+105}

\def\nicer{{NICER}}

\def\xspec{{\tt XSPEC}}

\begin{document}

\title{\vspace{-0.0cm}The 2022 Outburst of \igr: \\
Connecting the exotic \grs\ to standard black hole X-ray binaries }
\author{\vspace{-1.5cm}Jingyi~Wang\altaffilmark{1},
Erin~Kara\altaffilmark{1}, 
Javier~A.~Garc{\'\i}a \altaffilmark{2,3}, 
Diego~Altamirano\altaffilmark{4},
Tomaso~Belloni\altaffilmark{5}$^{,*}$,
James~F.~Steiner\altaffilmark{6},
Michiel van der Klis\altaffilmark{7}, 
Adam~Ingram\altaffilmark{8}, 
Guglielmo~Mastroserio\altaffilmark{9}, 
Riley~Connors\altaffilmark{10}, 
Matteo~Lucchini\altaffilmark{7},
Thomas~Dauser\altaffilmark{11}, 
Joseph~Neilsen\altaffilmark{10}, 
Collin~Lewin\altaffilmark{1},
Ron~A.~Remillard\altaffilmark{1},
Jeroen~Homan\altaffilmark{12}.
}
\affil{
\altaffilmark{1}MIT Kavli Institute for Astrophysics and Space
Research, MIT, 70 Vassar Street, Cambridge, MA 02139, USA\\
\altaffilmark{2}X‐ray Astrophysics Laboratory, NASA Goddard Space Flight Center, Greenbelt, MD 20771\\ 
\altaffilmark{3}Cahill Center for Astronomy and Astrophysics, California Institute of Technology, Pasadena, CA 91125, USA\\ 
\altaffilmark{4} School of Physics and Astronomy, University of Southampton, Highfield, Southampton, SO17 1BJ\\
\altaffilmark{5} INAF-Osservatorio Astronomico di Brera, via E. Bianchi 46, I-23807 Merate, Italy\\
\altaffilmark{6}Harvard-Smithsonian Center for Astrophysics, 60 Garden St., Cambridge, MA 02138, USA \\
\altaffilmark{7}Astronomical Institute, Anton Pannekoek, University of Amsterdam, Science Park 904, NL-1098 XH Amsterdam, Netherlands\\
\altaffilmark{8}School of Mathematics, Statistics and Physics, Newcastle University, Herschel Building, Newcastle upon Tyne, NE1 7RU, UK\\
\altaffilmark{9} INAF-Osservatorio Astronomico di Cagliari, via della Scienza 5, I-09047 Selargius (CA), Italy\\
\altaffilmark{10} Villanova University, Department of Physics, Villanova, PA 19085, USA\\
\altaffilmark{11}Remeis Observatory \& ECAP, Universit\"{a}t
Erlangen-N\"{u}rnberg, 96049 Bamberg, Germany\\ 
\altaffilmark{12} Eureka Scientiﬁc, Inc., 2452 Delmer Street, Oakland, CA 94602, USA\\
}

\begin{abstract}
\vspace*{-.3em}
While the {\color{black}standard} X-ray variability of black hole X-ray binaries (BHXBs) is stochastic and noisy, there are two {\color{black}known} BHXBs that exhibit exotic `heartbeat'-like variability in their light curves: \grs\ and \igr. {\color{black}In 2022}, \igr\ went into outburst for the first time in the NICER/NuSTAR era. These exquisite data allow us to simultaneously track the exotic variability and the corresponding spectral features with unprecedented detail. We find that {\color{black}as in typical BHXBs, the outburst began in the hard state, then the intermediate state, but then transitioned to an exotic soft state where we identify} two types of heartbeat-like variability (Class V and a new Class X). 
The flux-energy spectra show {\color{black}a} broad iron emission line due to relativistic reflection when there is no exotic variability, and absorption features from highly ionized iron when the source exhibits exotic variability. {\color{black}Whether absorption lines from highly ionized iron are detected in \igr\ is not determined by 
the spectral state alone, but rather is determined by the presence of exotic variability; in a soft spectral state, absorption lines are only detected along with exotic variability.} Our finding indicates that \igr\ can be seen as a bridge between the most peculiar BHXB \grs\ and `normal' BHXBs because it alternates between the conventional and exotic behavior of BHXBs. We discuss the physical nature of the absorbing material and exotic variability in light of this new legacy dataset.

\end{abstract}
\keywords{accretion, accretion disks 
--- black hole physics}

\section{Introduction}\label{intro}
\blfootnote{$^*${\color{black}We dedicate this paper to the late Tomaso Belloni, who contributed significantly to this paper before his untimely passing on 26 August 2023. Tomaso was a pioneer in the study of X-ray timing since his early days working on EXOSAT, and in particular, awakened the community to the beautiful puzzle that is GRS 1915. In this work, on GRS 1915's `little sister', IGR J17091, we build upon the legacy of a trailblazer in our field. We will miss him for his energy, his insights, his humor and his unwavering passion for science. Ad astra, Tomaso.}} BHXBs provide us with opportunities to study different accretion states in a single source on a human timescale. In a typical outburst of BHXBs, they rise from {\color{black}quiescence to a} \textit{hard state} where the X-ray emission is dominated by emission from the `corona' ({\color{black}the} hot plasma with temperature on the order of $100$~keV). Then, they make a rapid state transition {\color{black}usually on a time scale of days to weeks} (through what is known as the \textit{intermediate state}) into the \textit{soft state} where the disk emission dominates. Finally, they come back to the hard state and then {\color{black}recede again into} quiescence (see {\color{black}e.g., \citealp{2011BASI...39..409B}, and \citealp{2022arXiv220614410K} for a recent review). }{\color{black}Standard BHXBs show low-frequency quasi periodic oscillations \citep[LFQPOs; see the review][and references therein]{ingram2019review} in their power spectra. The LFQPOs in BHXBs are usually categorized with an {\color{black}A/B/C} classification scheme (see e.g., \citealp{2011MNRAS.418.2292M}). Type-C QPOs are strong ($\lesssim20\%$ rms) and narrow ($Q\gtrsim6$), and sit on top of a flat-top noise whose high-frequency break is close to the QPO frequency. They are seen commonly in the hard state and hard-intermediate state (HIMS). Type-B QPOs are seen in soft-intermediate state (SIMS), and they are narrow ($Q\gtrsim6$) but weaker compared to Type-C ($\lesssim5\%$ rms), found usually at 5--6~Hz and sometimes 1--3 Hz. They appear on top of weak red noise. 
Type-A QPOs are very rare, weak (a few percent rms), and broad ($Q\lesssim3$), and they are accompanied by very weak red noise.} 

\igr\ and \grs\ are extraordinary BHXBs because they are the only two {\color{black}known} BHXBs that {\color{black}exhibit} a variety of \textit{exotic variability} {\color{black}classes}, usually consisting of flares and dips that are highly structured and have high amplitudes (e.g., \citealp{2000A&A...355..271B,altamirano2011faint,2017MNRAS.468.4748C}). Depending on the characteristics of flares and dips, there are distinct variability classes observed: {\color{black}14} in \grs\ \citep{2000A&A...355..271B,2002MNRAS.331..745K,2005A&A...435..995H} and 9 in \igr\ \citep{2017MNRAS.468.4748C}. Out of the 9 classes, 7 classes of \igr\ resemble those in \grs, including the famous `heartbeat' variability mimicking an electrocardiogram, and the other 2 are unique to \igr. {\color{black}Because of the famous `heartbeat' class (\textit{Class IV} in \igr\ and \textit{Class $\rho$} in \grs), in this work, we refer to variabilities that are structured and repeated as `exotic' or `heartbeat-like'. It is also worth noting} that high-frequency QPOs (HFQPOs) are detected at the same frequency, 66~Hz, in \grs\ and \igr\ \citep{1997ApJ...482..993M,2012ApJ...747L...4A}. {\color{black}The} variability in \igr\ is generally faster than in the corresponding class in \grs\ \citep{altamirano2011faint,2017MNRAS.468.4748C}. 

{\color{black}\igr\ has had 8 outbursts in the past 30 years (see a summary in Section~2.2.26 in \citealp{tetarenko2016watchdog}). The outbursts in 1994, 1996, and 2001 were identified through an archival search after the first discovery of the source in 2003 \citep{2003ATel..149....1K}. In both the 2003 and 2007 outbursts, a transition from hard to soft state was found based on spectral and timing properties akin to typical BHXBs \citep{2006ApJ...643..376C, 2009ApJ...690.1621C}. The following 2011 outburst was the most extensively studied one, and this is when the heartbeat-like variability reminiscent of \grs\ was observed for the first time in this source (e.g., \citealp{altamirano2011faint}). The mass of the compact object or companion star in \igr\ is unknown and no parallax distance is available. }

{\color{black}On the other hand, \grs\ is a $12\pm2$~$M_\odot$ black hole accreting matter from a 0.8~$M_\odot$ K-giant companion in a wide 33.5-day orbit, and the parallax distance to it is $8.6^{+2.0}_{-1.6}$~kpc \citep{2001Natur.414..522G,2014ApJ...796....2R}. It is a peculiar BHXB as it remained in a persistent bright outburst for 26 years since its discovery in 1992 \citep{1992IAUC.5590....2C}, exhibiting a variety of exotic variability classes. In 2018, the source started to fade exponentially and settled in a faint (only a few percent of its previous flux) hard state in 2019 \citep{2018ATel11828....1N,2019ATel12742....1H}. }

While {\color{black}the X-ray} variability of BHXB lightcurves is attributed to stochastic and noisy coronal variability, the exotic variability is generally thought to be due to limit-cycle {\color{black}instabilities at} the inner accretion disk. The most {\color{black}common hypothesis} for the origin of such instability is the radiation pressure instability \citep{2000ApJ...535..798N,2000ApJ...542L..33J,done2004grs,2011ApJ...737...69N}.
The radiation pressure instability requires the source to accrete at a high Eddington ratio (e.g., $>26\%$~$L_{\rm Edd}$ in \citealp{2000ApJ...535..798N}), which is plausible for \grs\ as it accretes at above a few tens percent of its Eddington limit and even super-Eddington rates \citep{done2004grs,2004ARA&A..42..317F,2011ApJ...737...69N}. However, {\color{black}this hypothesis has been questioned} since similar exotic variability was discovered in \igr. With {\color{black}a flux that is $\sim20$--30 times lower in} \igr\ compared to \grs, a high-Eddington-accretion scenario means \igr\ either harbors the lowest mass black hole known ($<3$~$M_\odot$ if $d<17$~kp) or {\color{black}it is very distant, or the compact object in \igr\ is a neutron star \citep{altamirano2011faint}.}

The disk-wind-jet connection in both \grs\ and \igr\ could shed light on the nature of the exotic variability. In the bright 2011 outburst of \igr, an absorption line at $6.91\pm0.01$~keV was revealed in one Chandra {\color{black}High Energy Transmission Grating (HETG)} spectrum, corresponding to an extreme outflow velocity of $0.03c$ if associated with a blueshifted Fe XXV line \citep{king2012extreme}. Later, \citet{janiuk2015interplay} noted that in the 2 Chandra observations in 2011, the presence of absorption lines and heartbeat variability were anti-correlated. {\color{black}These authors} proposed that a disk wind might stabilize the disk and suppress the heartbeat pattern. However, {\color{black}\citet{reis2012igr} found contradicting evidence with the discovery of} a tentative absorption line at 7.1~keV {\color{black}coincident} with the heartbeat variability using XMM-Newton {\color{black}EPIC-pn} data. 

\begin{figure*}
\centering
\includegraphics[width=0.8\linewidth]{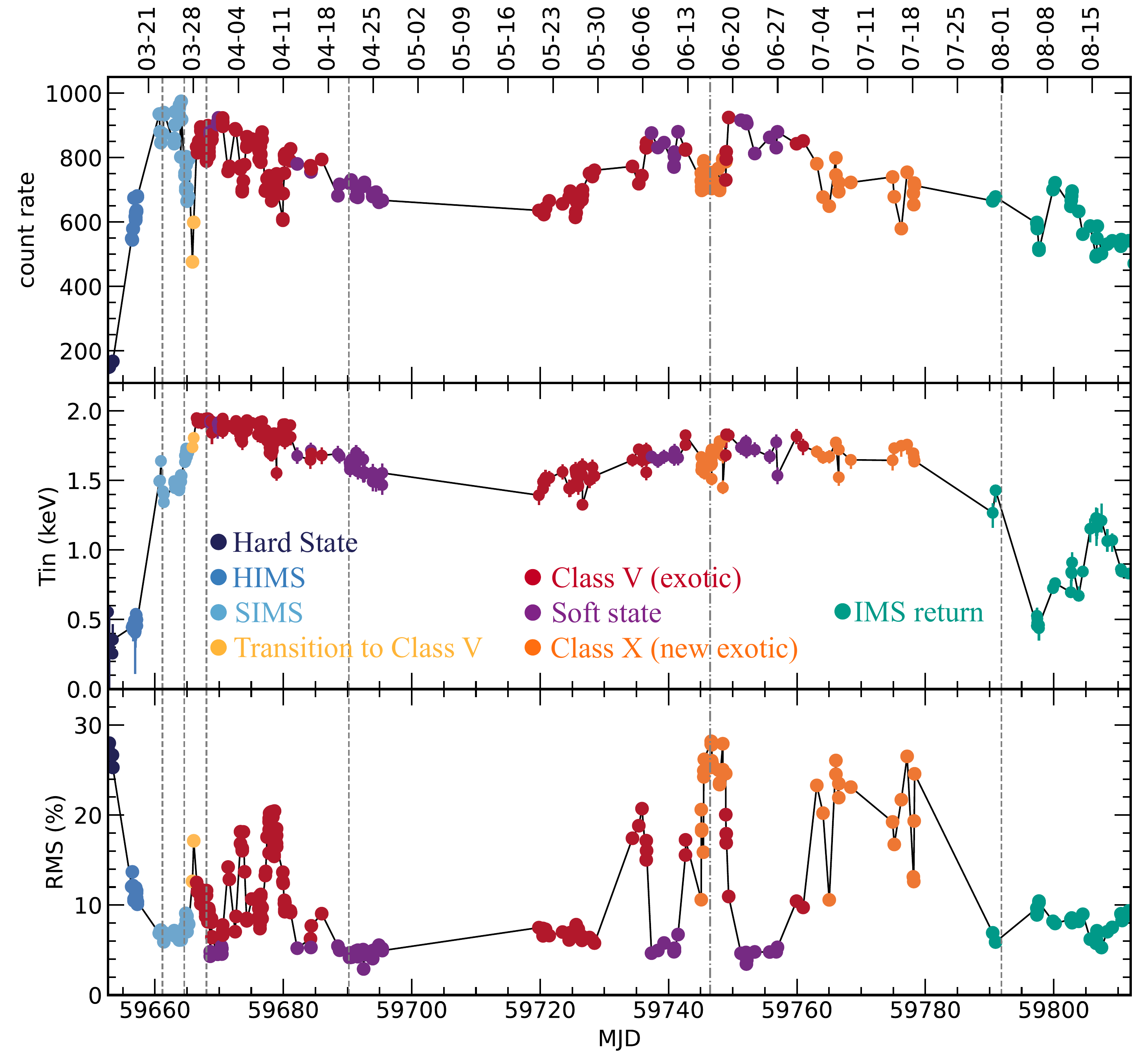}
\caption{The time evolution of NICER count rate (0.3--12~keV, normalized for 52~FPMs), the fitted disk temperature with a baseline model (see Section~\ref{sect:auto_fit}), and the fractional rms (0.01--10~Hz in 1--10~keV). {\color{black}There are 305 data points, each representing a 500~s NICER segment used in both spectral and timing analysis. The color coding is based on the state identification in Section~\ref{sect:preliminary_states}. The {\color{black}gray} lines indicate when the 6 NuSTAR observations take place{\color{black}, and the dash-dotted line marks June 16 during which the Chandra/HETG and the fifth NuSTAR observations take place} (see Table~\ref{tab:obs}). Besides MJD, the calendar dates are shown on the top x-axis.}}
\label{fig:mjd}
\end{figure*}

\begin{table*}
\begin{center}
\caption{The observation catalog. \label{tab:obs}}
\footnotesize
\begin{tabular}{@{\extracolsep{8pt}}ccccccccc@{}}\hline \hline
State/ & \multicolumn{4}{c}{NICER} & \multicolumn{4}{c}{NuSTAR} \\\cline{2-5} \cline{6-9} 
Variability Class& Date & Exp.(ks) & Counts s$^{-1}$ & rms (\%) & ObsID & Date & Exp.(ks) & Counts s$^{-1}$ \\
\hline
\textit{Hard State} & 03/14--03/16 & 2.0 & 140 & 27 & $\dots$ & $\dots$ & $\dots$ & $\dots$ \\
\textit{HIMS} & 03/18--03/19 & 10.0 & 562 & 12 & $\dots$ & $\dots$ & $\dots$ & $\dots$\\
\multirow{2}{*}{\textit{SIMS}} & \multirow{2}{*}{03/22--03/27} & \multirow{2}{*}{13.0} & \multirow{2}{*}{770} & \multirow{2}{*}{8} & 80702315002 & 03/23 & 11.3 & 87 \\
& &  &  & & 80702315004 & 03/26 & 16.5 & 71 \\
\textit{Transition to Class V} & 03/27--03/28 & 2.2 & 556 & 16 & $\dots$ & $\dots$ & $\dots$ & $\dots$\\
\textit{Class V} (exotic) & 03/28--06/30 & {\color{black}67.5} & {\color{black}721} & {\color{black}12} & 80702315006 & 03/29 & 11.9 & 94 \\ 
\textit{Soft State} & 03/30--06/26 & {\color{black}25.0} & {\color{black}735} & {\color{black}5} & 80802321002 & 04/21 & 15.1 & 76 \\
\textit{Class X} (new exotic) & 06/15--07/18 & 19.5 & 648 & 24 & 80802321003 & 06/16 & 16.1 & 69 \\
\textit{IMS Return} & 07/30--08/21 & 14.5 & 504 & 8 & 80802321005 & 07/31 & 13.9 & 61 \\ 
\hline
\hline
\end{tabular}
\\
\raggedright{\textbf{Notes.} \\
The source makes excursions between \textit{Class V}, \textit{Class X}, and the \textit{Soft State} from March 28 to July 18, so the dates listed for these 3 classes are the initial start date and final end date. The exposure times of NICER are the total exposure time of the 500s segments used in {\color{black}this work}, except for that in the \textit{Transition to Class V}, we combine all the available data for the flux-energy spectrum to increase the signal-to-noise ratio. {\color{black}Otherwise, there would be only 2 segments of 500s in the \textit{Transition to Class V}.} The NICER count rate and rms are in 1--10~keV, and the NuSTAR count rate is in 3--78~keV.
The Chandra/HETG observation (ObsID 26435) has an exposure of 30~ks and was taken on June 16 in \textit{Class X}. }
\end{center}
\end{table*}

In this paper, we present the spectral-timing analysis of \igr\ in its 2022 outburst using our observing campaign with the Neutron Star Interior Composition Interior Explorer (NICER; \citealp{gendreau2016neutron}), Nuclear Spectroscopic Telescope Array (NuSTAR; \citealp{harrison2013nuclear}), and Chandra/HETG \citep{2005PASP..117.1144C}. {\color{black}During this campaign, the source exhibited complex phenomenology, which we attempt to classify into different states based on the spectral and timing properties of the source. After a brief description of the observations and data reduction in Section~\ref{obs}, we begin Section~\ref{sect:tools} by first describing the methods we use to classify each state. Namely, we identify the different states by (1) the spectral shape, and (2) the shape of the light curves. 
After identifying the different states in Section~\ref{sect:preliminary_states}, we perform detailed power spectral (Section~\ref{sect:psd}) and flux-energy spectral analysis (Section~\ref{sect:detailed_spec}) of each state, to understand how the physics of the accretion flow changes in each state. We summarize the key properties of each state in Section~\ref{sect:states}. Finally, we discuss and interpret our findings in Section~\ref{discussion}.}

\section{Observations and data reduction}\label{obs}
\subsection{Observations}
After its last outburst in 2016, \igr\ entered a new outburst in March 2022 \citep{miller2022nicer}. {\color{black}When this outburst began, we triggered our NICER and NuSTAR GO Program (PI: J. Wang). Here we analyze all 136 NICER observations taken at a near-daily cadence from 2022 March 27 to Aug 21, as well as 6 NuSTAR observations taken over this same epoch (see the observation catalog in Table~\ref{tab:obs}). We also requested (by Directors Discretionary Time) one Chandra/HETG observation during this campaign{\color{black}, and this observation took place on June 16, which was simultaneous to the fifth NuSTAR observation}.}
The time evolution of the count rate, fitted disk temperature (see Section~\ref{sect:auto_fit}), and the fractional rms are shown in Fig.~\ref{fig:mjd}.

\subsection{Data reduction}
\subsubsection{NICER}
{\color{black}We process the \nicer\ data} with the data-analysis software NICERDAS version v2020-04-23\_V007a, and energy scale (gain) release `optmv10'. We use the following filtering criteria: the pointing offset is less than $60\arcsec$, the pointing direction is more than $30^\circ$ away from the bright Earth limb, and more than $15^\circ$ away from the dark Earth limb, and the spacecraft is outside the South Atlantic Anomaly (SAA).  Data are required to be collected at either a sun-angle $>60^\circ$ or else collected in shadow (as indicated by the `sunshine' flag).  We filter out commonly-noisy detectors FPMs \#14, 34, and 54. In addition, we flag any `hot detectors' in which X-ray or undershoot rates {\color{black}(detector resets triggered by accumulated charge)} are far out of line with the others ($\sim 10 \sigma$) and exclude those detectors for the GTI in question.  We select events that are not flagged as `overshoot' {\color{black}(typically caused by a charged particle passing through the detector and depositing energy)} or `undershoot' resets (EVENT\_FLAGS=bxxxx00), or forced triggers (EVENT\_FLAGS=bx1x000), and require an event trigger on the slow chain {\color{black}which is optimized for measuring the energy of the event (i.e., excluding fast-chain-only events where the fast chain is optimized for more precise timing).} A `trumpet' filter on the `PI-ratio' is also applied to remove particle events from the detector periphery \citep{bogdanov2019constraining}. The resulting cleaned events are barycenter corrected using the \texttt{FTOOL} \texttt{barycorr}. The background spectrum is estimated using the 3C50 background model \citep{2022AJ....163..130R}. GTIs with overshoot rate $>2$~FPM$^{-1}$~s$^{-1}$ are excluded to avoid unreliable background estimation. We use the RMF version `rmf6s' and ARF version `consim135p', which are both a part of the CALDB xti20200722. {\color{black}We also add 1\% systematics to the NICER spectra at energies below 3~keV to account for the effects of calibration uncertainties. The fitted energy range in the flux-energy spectral analysis is 1--10~keV. }

\subsubsection{NuSTAR}
The NuSTAR data are reduced using data analysis software (NUSTARDAS) 2.1.2 and CALDB v20220802. Due to elevated background rates around the SAA, the data are processed using \texttt{nupipeline} with `saamode=strict' and `tentacle=yes'. The source spectra and lightcurves are extracted from circular regions with a radius of 100$''$, and the background is from off-source regions of the same size. We also note that stray light contamination is present in the field of view of focal plane module FPMB in several observations, leading to increased background. Both NICER and NuSTAR spectra are then oversampled in energy resolution by a factor of 3 and are binned with a minimum count of 25 per channel. {\color{black}For spectral analysis, the fitted energy range is 3--78~keV for NuSTAR observations 1, 2, and 6, and 3--20~keV for NuSTAR observations 3--5 whose spectra are soft, and therefore background dominates at energies above $\sim$20~keV. }

\begin{figure}
\centering
\includegraphics[width=1.\linewidth]{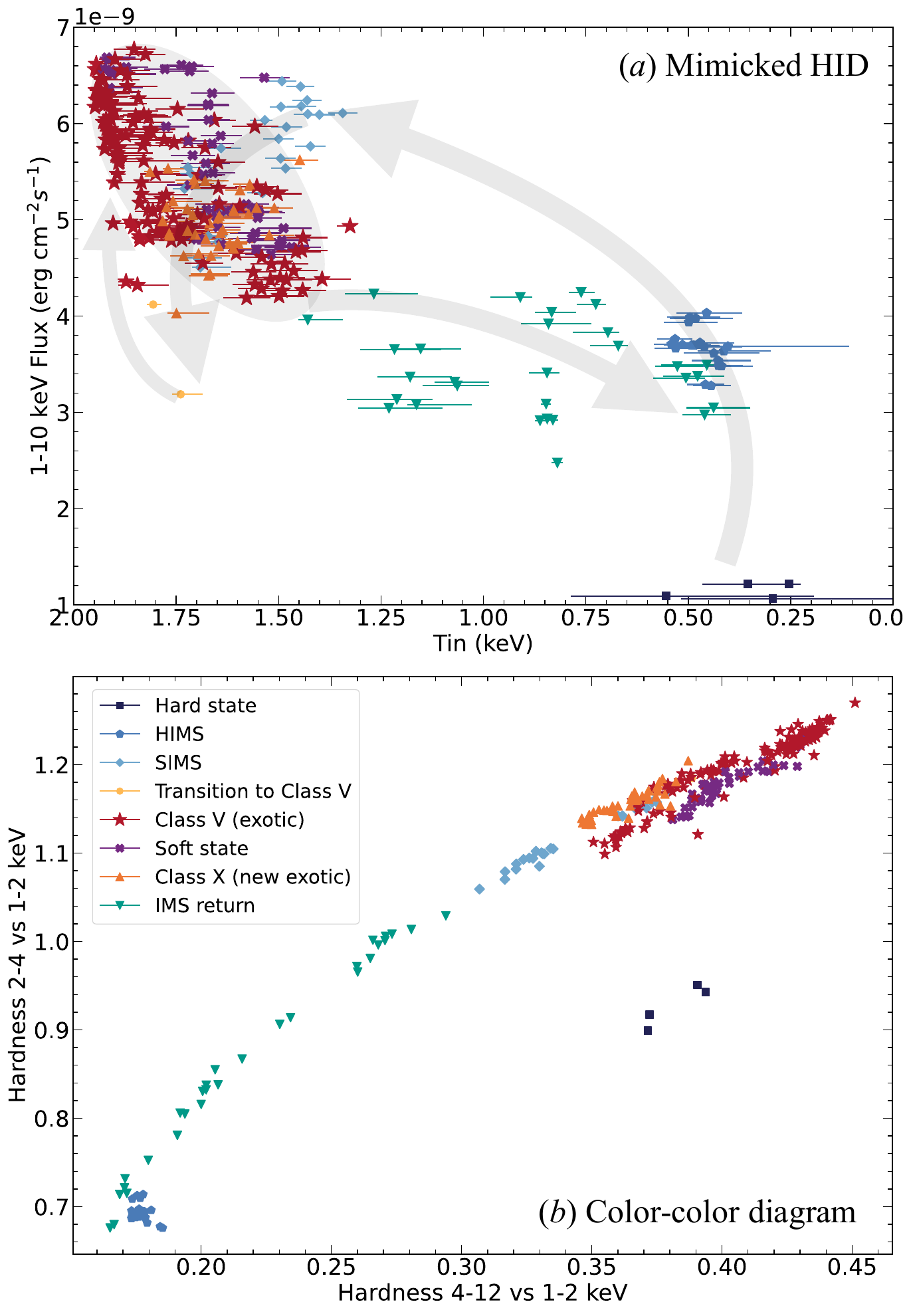}
\caption{(\textit{a}) The mimicked HID to show the spectral state evolution. Both the total flux (1--10~keV) and the disk temperature are measured using the baseline model on the 305 NICER 500s segments (Section~\ref{sect:auto_fit}). The gray arrows indicate the evolution in time. (\textit{b}) The NICER color-color diagram, where the colors are defined as the count rate ratios between 4--12 and {\color{black}1--2 keV}, and 2--4 and 1--2~keV. The color coding is based on the state identification in Section~\ref{sect:preliminary_states}. }
\label{fig:hid_cc}
\end{figure}

\begin{figure*}
\centering
\includegraphics[width=1.0\linewidth]{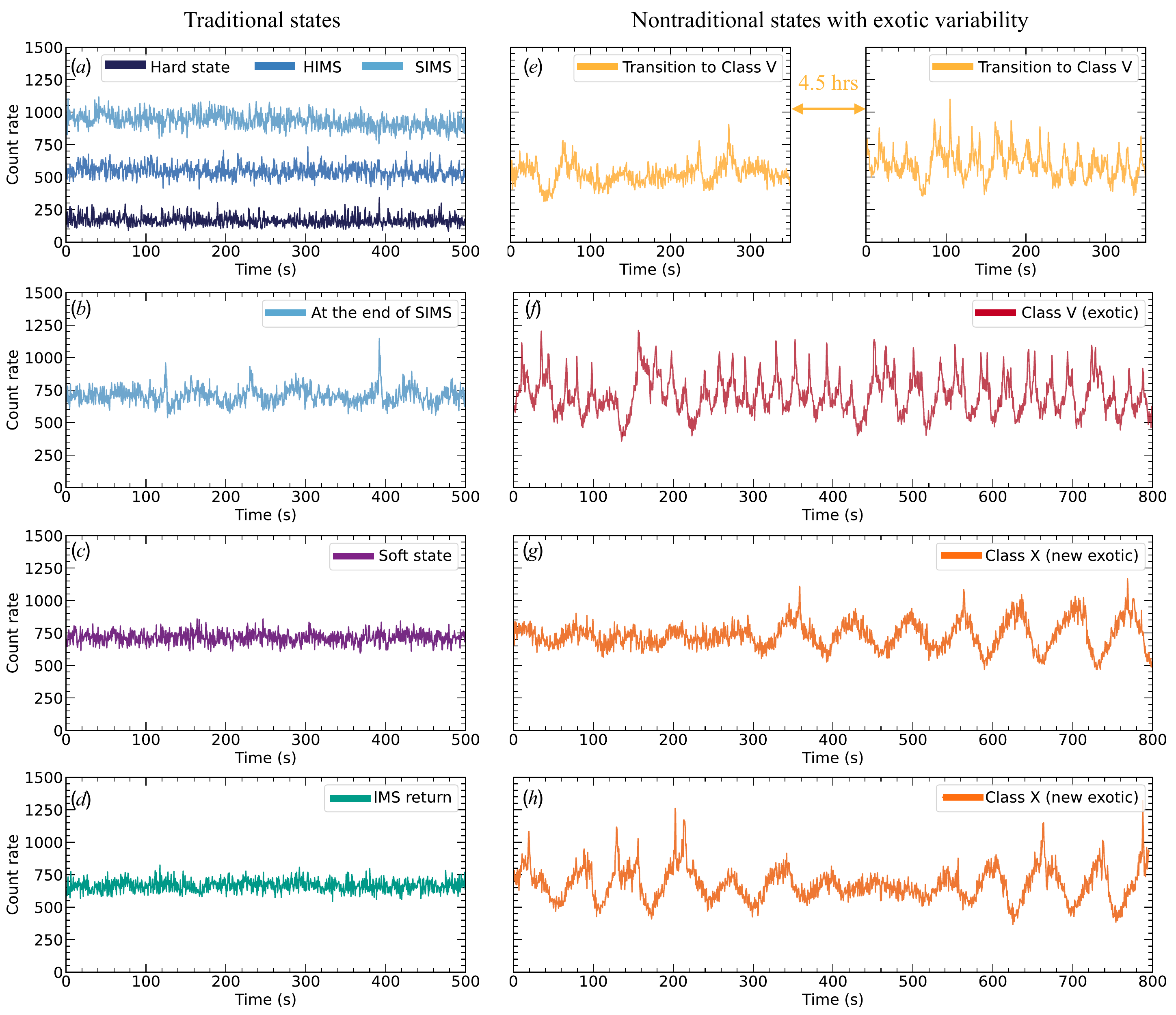}
\caption{Representative NICER lightcurves in each state or variability class. 
{\color{black}The outburst started in the \textit{Hard State}, \textit{HIMS}, and then \textit{SIMS} (panel \textit{a}), when the variability was stochastic; at the end of \textit{SIMS}, exotic variability started to show up (panel \textit{b}; see also the PSD in Fig.~\ref{fig:psd}b). During the \textit{Transition to Class V} (panel \textit{e}), the heartbeat-like exotic variability developed during $4.5$~hours (see more lightcurves in Fig.~\ref{fig:lc_transition}). The source then transitioned back and forth between the \textit{Soft State} (panel \textit{c}), \textit{Class V} (exotic; panel \textit{f}), and \textit{Class X} (new exotic; panels \textit{g}--\textit{h}). Finally, the source went back into the \textit{IMS Return} (panel \textit{d}) when the variability became stochastic again.}
The count rate is measured in 0.3--12~keV with NICER normalized for 52 FPMs, with time bins of 0.5~s. {\color{black}The NICER ObsIDs are 5202630102 (\textit{Hard State}), 4618020101 (\textit{HIMS}), {\color{black}4618020202} (\textit{SIMS}), 4618020402 (at the end of \textit{SIMS}), {\color{black}5618010403} (\textit{Soft State}), 5618011401 (\textit{IMS Return}), 5202630108 and 5202630109 (\textit{Transition to Class V}), 5202630116 (\textit{Class V}), 5618010802 and 5618011202 (\textit{Class X}).}  }
\label{fig:lc}
\end{figure*}

\subsubsection{Chandra}
{\color{black}We reprocess the Chandra/HETG data (ObsID 26435)} using CIAO v4.14 and CALDB v4.9.7. We follow the standard data reduction
process for the grating data and decrease the width of the masks on the grating arms used to extract the spectra from the default of 35 to 18 pixels. This decreases the overlap between the HEG and MEG arms and thus allows us to extend our analysis to higher
energies. First-, second-, and third-order spectra were extracted from the observation, and the positive and negative spectra for each order were combined to increase signal-to-noise ratio with \texttt{combine\_grating\_spectra}. 

\vspace{0.3cm}
All the uncertainties quoted in this paper are for a 90\% confidence range unless otherwise stated. {\color{black}We use XSPEC 12.12.1 \citep{arn96} for all the spectral fits.} In all of the fits, we use the \textit{wilm} set of abundances \citep{wilms2000}, \textit{vern} photoelectric cross sections \citep{verner1996atomic}, and $\chi^2$ fit statistics. 

\section{Methodology for Identification of States}
\label{sect:tools}

{\color{black} With the nearly daily cadence of our NICER observations, we are able to track the source extensively, as it evolved in its spectral and timing characteristics. The phenomenology of \igr\ is particularly complex. In this section, we attempt to bring order to this complexity by categorizing phenomenology and comparing it to previous observations. Here, we present different methods that we use to describe the phenomenology in each observation, namely, their spectral shape (Section~\ref{sect:auto_fit}), their light curve shapes (Section~\ref{sect:lc}), and summarize our finding (Section~\ref{sect:preliminary_states}). We will then use these state identifications and names in the remainder of the paper. }



\subsection{The Broadband Continuum Shape} \label{sect:auto_fit}

{\color{black}To decipher the spectral states, we begin by identifying the dominant spectral component in each observation. In an automated way, we fit the flux-energy spectra of all the 305 NICER segments of the length of 500~s (i.e., continuous 500s intervals), making a total exposure time of 152.5~ks. The baseline model used includes} the multi-color disk emission (\texttt{diskbb}) and a Comptonization component (\texttt{nthcomp}). We use \texttt{cflux} to calculate the flux contribution from each component. The \xspec\ syntax of the model, therefore, is \texttt{TBabs(cflux*diskbb + cflux*nthComp)}. 

The time evolution of the fitted disk temperature is shown in Fig.~\ref{fig:mjd}. At the beginning of the outburst, the disk temperature is low ($kT\sim 0.5$~keV), and rises to 1.5--2~keV as the luminosity increases. We attempt to place these observations in the conventional hardness-intensity diagram (HID) in order to cleanly identify hard and soft states, but because of these very high disk temperatures and the high galactic absorption column ($N_{\rm H}>10^{22}$~cm$^{-2}$), in some observations more thermal-dominated spectra actually led to larger hardness ratios (e.g., see either hardness ratio in the color-color diagram in Fig.~\ref{fig:hid_cc}b). In other words, the conventional phenomenological HID fails to capture the corona-dominated states versus the thermal-dominated states. To overcome this, we plot the fitted disk temperature as a proxy for the spectral hardness\footnote{{\color{black}We also tried to use the disk fraction (disk flux divided by the total flux) as the x-axis in the mimicked HID, and found a consistent pattern tracked by the source, but the contribution from the Compton component can be very difficult to be constrained with short NICER segments.}}. The resulting `mimicked' HID is shown in Fig.~\ref{fig:hid_cc}(a). With this approach, we can map the evolution of \igr\ in its 2022 outburst to more typical BHXBs. Following the classical pattern, \igr\ started the outburst in the \textit{Hard State} (i.e. at low disk temperature, on the bottom-right side of the mimicked HID), rose in flux, and transitioned to the higher temperatures, corresponding to the \textit{Hard Intermediate State} (\textit{HIMS}), \textit{Soft Intermediate State} (\textit{SIMS}), and the \textit{Soft State}, and eventually went back towards the hard state at a lower flux than the hard-to-soft state transition. This is akin to the hysteresis pattern seen in the HIDs of typical BHXBs. 


\subsection{{\color{black}Light Curves}} \label{sect:lc}

{\color{black}Fig.~\ref{fig:lc} shows representative NICER lightcurves discovered during our observing campaign. The shapes of the light curves vary dramatically, and these shapes can be broadly characterized into different states (see also Fig.~\ref{fig:mjd} for when each state was observed).}


{\color{black}At the beginning of the outburst (from March 14 to 17), \igr\ showed stochastic variability (Fig.~\ref{fig:lc}, Panel a, navy curve), which coincided with the hardest spectra, when the disk temperature was lowest, at $\sim 0.5$~keV (see also \citealp{miller2022nicer}). On March 18, the source flux began to increase, akin to the spectral \textit{Hard Intermediate State} (\textit{HIMS}; March 18 to 19) and \textit{Soft Intermediate States} (\textit{SIMS}; March 22 to 27; \citealt{wang2022nicer_a}), but the variability remained stochastic (Panel a, cyan and green curves). By March 26, exotic variability started to develop in the lightcurves (Panels b and e, coined {\it `At the end of SIMS'} and the {\it `Transition to Class V'}; \citealt{wang2022nicer_b}), and by March 28, when the disk temperature was near its highest values, between 1.5--2~keV (akin to a soft spectral state), the light curves showed very clear, structured variability (Panels f--h). The exotic variability seen in Panel f is reminiscent of {\it `Class V'} variability identified in \citet{2017MNRAS.468.4748C}, while the near sinusoidal variability seen in Panels g and h do not resemble any previously identified classes. Therefore, we coin this new variability, as `\textit{Class X}' (more details below). From March 28 to July 18, \igr\ made excursions between variability \textit{Class V}, \textit{Class X}, and a more `traditional' \textit{Soft State} (Panel c) that shows very little variability altogether.}

\begin{figure}
\centering
\includegraphics[width=1.\linewidth]{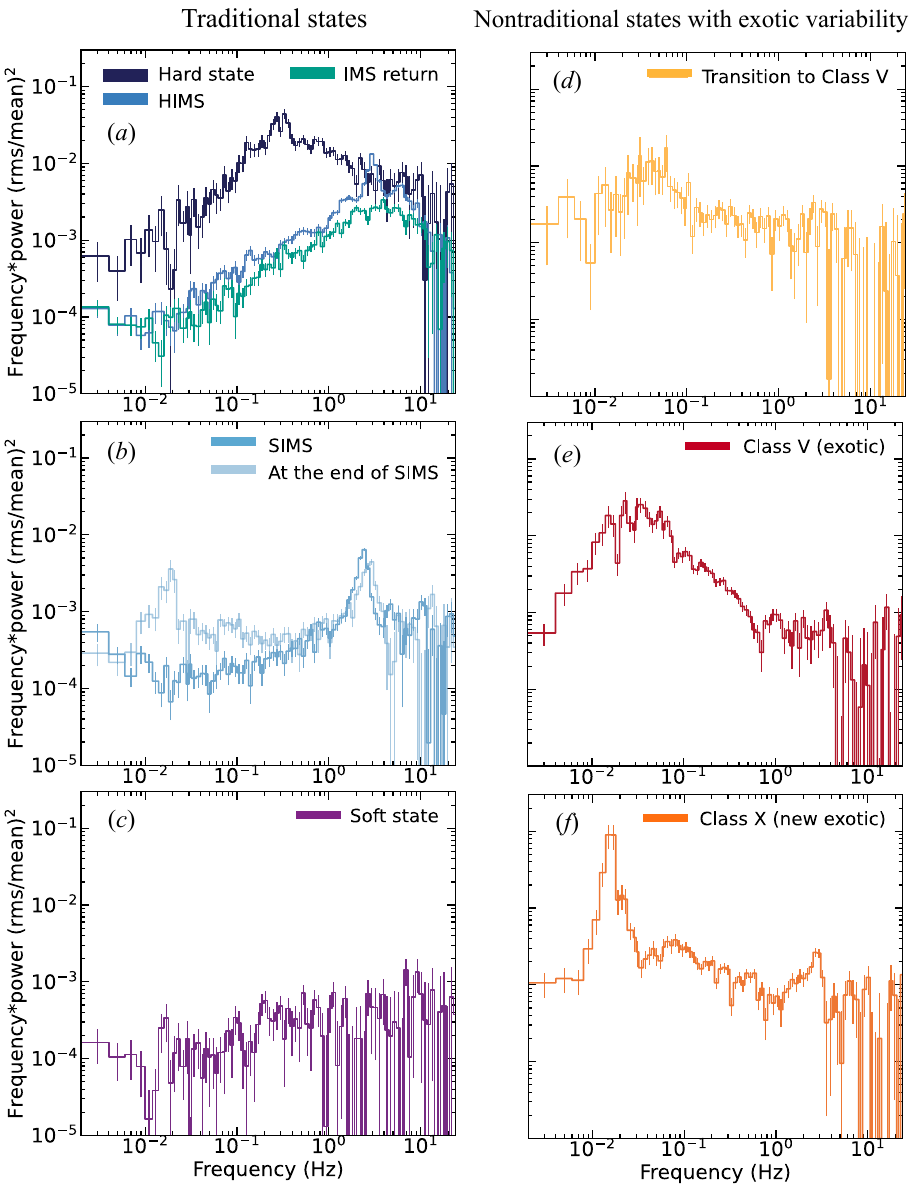}
\caption{{\color{black}Representative PSDs in each state or variability class, which are averaged over $\gtrsim10$ segments of the length of 500 seconds to increase the signal-to-noise ratio. The logarithmic frequency rebinning factor is 0.1.} }
\label{fig:psd}
\end{figure}

\begin{figure*}
\centering
\includegraphics[width=1.\linewidth]{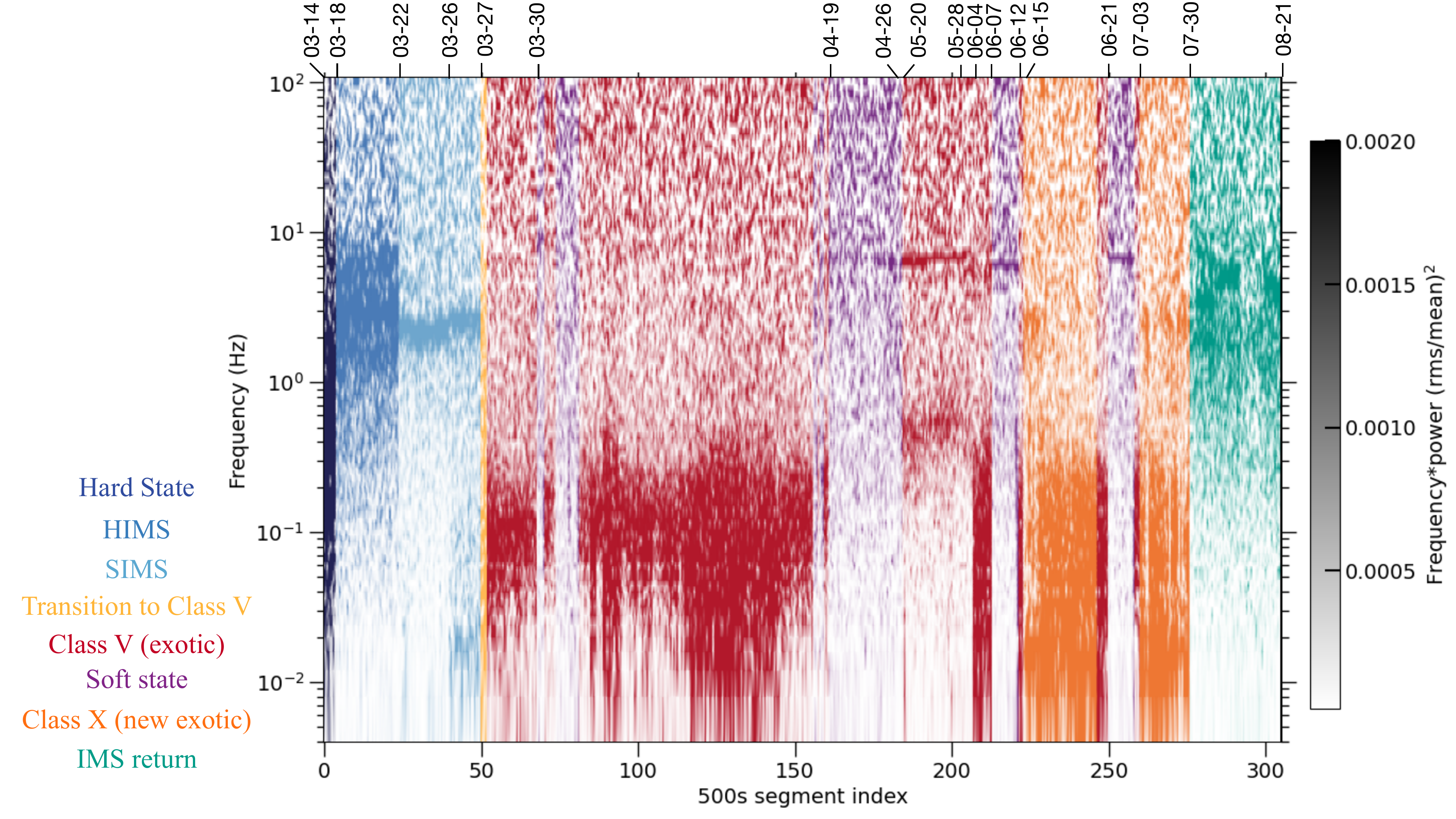}
\caption{The dynamical power spectrum using {\color{black}all the 305 NICER segments of length of 500~s} in 1--10~keV, color-coded based on the accretion state identification (see Section~\ref{sect:preliminary_states} for more details). {\color{black}For a given state/color, the darker shade corresponds to higher variability power, and we only show the grey scale for clarity. The x-axis is the index of the 500~s segments, and we also show the transitional dates on the top x-axis.}}
\label{fig:dpsd}
\end{figure*}

{\color{black}Here we describe the exotic variability classes in more detail. The structured, exotic variability began gradually at the end of the \textit{SIMS}, as sharp flares began arising on top of the stochastic variability (Panel b). 
{\color{black}Then on March 27, the lightcurves started to become more variable with different characteristics than in \textit{SIMS}. The NICER lightcurves from March 26 to March 28 are shown in Fig.~\ref{fig:lc_transition}. Comparing the lightcurves at March 27 06:10 and 18:49 UTC, the average NICER count rate decreased from $\sim800$ to $\sim600$ counts/sec. Later that day, the average count rate decreased even further to $\sim500$~counts/sec. Then, within $4.5$~hours}, the source went from demonstrating largely stochastic variability to showing distinct and highly structured exotic variability pattern, having firmly transitioned to \textit{Class V} variability (Panel~e). }

{\color{black}\textit{Class V} lightcurves (Panel f) are characterized as having repeated, sharp, high amplitude flares, although the period of those flares drifts even within a 500s segment. Each major flare's apex can be singly-peaked or multi-pronged, the latter of which we refer to as `mini-flares'.}

{\color{black}The lightcurves in the new \textit{Class X} (Panel g--h) show nearly sinusoidal variations. They are distinguished from \textit{Class V} by the larger amplitudes (mean rms is 24\% compared to 14\%), uniformity (see the power spectral density in Section~\ref{sect:psd}), and symmetry of the flares. Sometimes there can also be additional mini-flares at the peaks of major flares. In some observations, the \textit{Class X} appeared to nearly vanish, but then reappear a few hundred seconds later (Panel h).}

{\color{black}Finally, at the end of our campaign (Aug 21), we observed the source transition back to lower disk temperatures ($\sim 1$~keV), akin to a traditional Intermediate State, and also the stochastic variability re-emerged (Panel d). We term this state as \textit{Intermediate State return} (or \textit{IMS Return}). We note that while our observations stop on Aug 21, after this date, \igr\ was found to exhibit exotic variability once again. The analysis of this later behavior will be published in future work. }

\subsection{Class Identification} \label{sect:preliminary_states}

{\color{black}From our analysis of the broadband spectral shape (Fig.~\ref{fig:hid_cc}) we conclude that this remarkable source transitions between spectral and timing characteristics of typical BHXBs (e.g. the \textit{Hard State}, \textit{HIMS}, \textit{SIMS}, \textit{Soft State} and \textit{IMS Return}). Then, the shape of the lightcurves (Fig.~\ref{fig:lc}) revealed that there was a phase of \textit{Transition to Class V}, and that sometimes in the soft state, instead of showing very little variability (as in most BHXBs), \igr\ can demonstrate exotic (structured and repeated) variability. Therefore, we also identified exotic variability classes \textit{Class V} and \textit{Class X}.}
In this way, \igr\ can be seen as a bridge between the more typical BHXBs and \grs\, with its famously complex and exotic variability. In the next section, we delve further into the spectral and timing properties of each of these identified states.

As a note to the reader: for the remainder of this paper, we use {\color{black}green/blue/purple} colors for observations in the more typical/stochastically varying states (e.g., the \textit{Hard State}, \textit{HIMS}, \textit{SIMS}, \textit{Soft State}, and \textit{IMS return}), while the exotic variability states (\textit{Transition to Class V}, \textit{Class V} and \textit{Class X}) are shown in {\color{black}red/orange/yellow}.

\section{Results} \label{sect:results}

\subsection{{\color{black}Power Spectra and Dynamical Power Spectrum}} \label{sect:psd}

{\color{black}To quantitatively investigate the characteristics of the lightcurves, we compute power spectral densities (PSDs) of all the 305 NICER segments of the length of 500~s. We use the `rms-squared' normalization \citep{1990A&A...227L..33B}, and the Nyquist frequency is 500~Hz. Representative PSDs from each of our identified states corresponding to the lightcurves in Fig.~\ref{fig:lc} are shown in Fig.~\ref{fig:psd}. 
The PSDs in Fig.~\ref{fig:psd} were computed by averaging $\gtrsim10$ segments of the length of 500~s, and are binned geometrically in frequency, i.e., from frequency $\nu$ to $(1+f)\nu$, where $f$ is called the f factor (see Section~2.2 in \citealt{uttley2014x} for more details). We choose an f factor of 0.1 to measure the characteristic frequencies more precisely.}

As \igr\ can evolve very {\color{black}quickly}, on a timescale of a day, {\color{black}and so} we also compute the dynamical power spectral density (DPSD) to show the evolution of PSDs over time (Fig.~\ref{fig:dpsd}). {\color{black}The DPSD can be regarded as a matrix of the PSD of each 500~s segment. The DPSD has been color-coded by their identified state, where for a given state/color, the darker shade corresponds to higher variability power. The DPSD clearly reveals the exotic variability, as a peak in the power at around 0.1~Hz, but one can also see other characteristic frequencies popping out.}

\begin{figure}
\centering
\includegraphics[width=1.\linewidth]{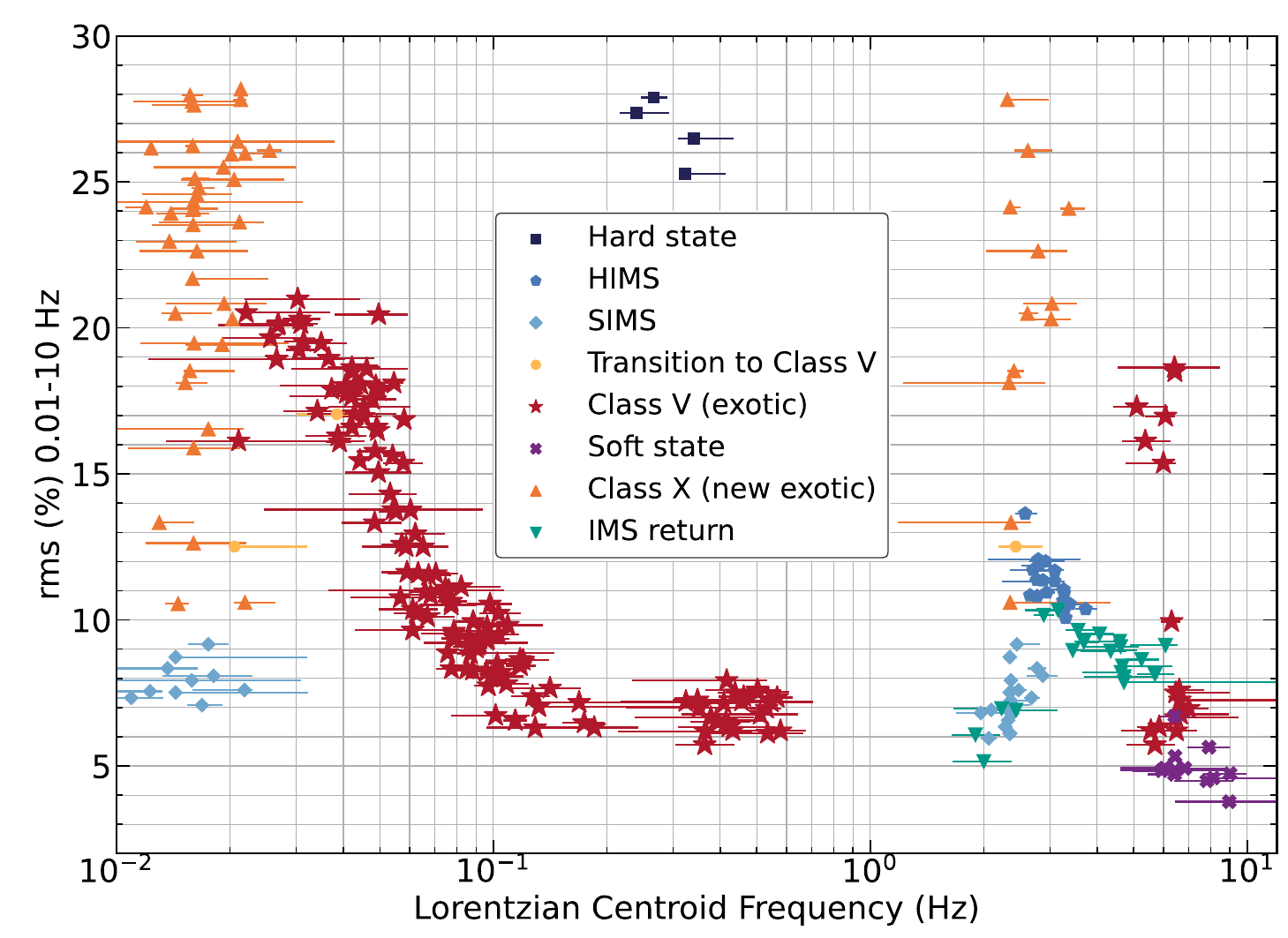}
\caption{The fitted characteristic frequencies (centroid frequencies of Lorentzians) in the {\color{black}single-segment} PSDs versus the fractional rms (0.01 to 10~Hz). The NICER energy band used is 1--10~keV. 
{\color{black}The data points in the \textit{Soft State} at 5--8~Hz correspond to a highly coherent QPO (see Wang et al., 2024b).} }
\label{fig:rms_f0}
\end{figure}

{\color{black}To investigate these characteristic frequencies further,} we then fit all the 305 raw (Poisson noise included) single-segment PSDs with multiple Lorentzian components and a constant for the Poisson noise. The Lorentzians {\color{black}of varying widths describe} both the broadband noise and the narrower components (including the {\color{black}`normal'} QPOs and the ones caused by exotic variability). We show the characteristic centroid frequencies of the narrower Lorentzian components versus the fractional rms in Fig.~\ref{fig:rms_f0}. {\color{black}Below, we describe some of the PSDs in Fig.~\ref{fig:psd} and characteristic timescales for each state and compare to typical BHXBs.}

In the \textit{Hard State} {\color{black}(Panel a, navy curve)}, we detect a QPO at $\sim0.3$~Hz with a Q factor $\sim6$ and a fractional rms $\sim13\%$. The QPO is accompanied by a flat-top noise with both low and high-frequency breaks. The high-frequency break is at a similar frequency to the QPO frequency. These characteristics are consistent with a Type-C QPO in a typical hard state. 

In the \textit{HIMS} {\color{black}(Panel a, cyan)}, both the frequencies of the QPO and the low-frequency break of the flat-top noise increase {\color{black}compared to the \textit{Hard State}}. The QPO is still narrow with a Q factor $\sim6$, and its fractional rms is $\sim6\%$. The QPO frequency is in the range of 2.5 to 5~Hz, and it also anticorrelates with the fractional rms (Fig.~\ref{fig:rms_f0}), which is a characteristic of Type-C QPOs in normal BHXBs (e.g., see \citealp{2011MNRAS.418.2292M}).

In the \textit{SIMS} {\color{black}(Panel a, green)}, a narrow and prominent QPO is always present at 2 to 3~Hz, with a weak power-law noise. No clear correlation between the QPO frequency and the rms can be seen (Fig.~\ref{fig:rms_f0}). The Q factor is $\sim5$ and the rms is $\sim5\%$. These features are consistent with a Type-B QPO in the traditional SIMS.

As found in Section~\ref{sect:lc}, \igr\ transitioned gradually from the \textit{SIMS} to the exotic \textit{Class V}, and the PSD in the \textit{Transition to Class V} has an increase of power generally across frequencies from 0.002 to $\sim10$~Hz (Panel~d). {\color{black}At the end of the \textit{SIMS},} because of the consistent Lorentzian centroid frequencies in the segment-based PSDs, we are able to fit the averaged PSD (Panel~b) using the multi-Lorentzian model. We measure Type-B QPO frequency at $\nu_1=2.70\pm0.06$~Hz, and an additional peak at $\nu_2=0.016^{+0.002}_{-0.001}$~Hz\footnote{{\color{black}We note that for the standard error on the PSD, we use the formula appropriate for a large number of samples ($KM\gtrsim20$): $\Delta P(\nu_j)=P(\nu_j)/\sqrt{KM}$ where $\nu_j$ is the frequency bin, and the PSD is averaged over $M$ segments and $K$ frequencies in bin $j$ \citep{uttley2014x}. For one single 500~s segment, the errors on the PSD do not approach Gaussian at frequencies as low as $\sim0.016$~Hz. Therefore we average over all 10 segments at the end of \textit{SIMS} to measure the centroid frequency. }}. The additional peak is caused by modulations occurring with a period of $1/\nu_2=62^{+4}_{-8}$~seconds (corresponding lightcurve in Fig.~\ref{fig:lc}b).

In \textit{Class V} {\color{black}(Panel~e)}, the lack of regularity in the flare period produces a PSD that can be fitted with a broad Lorentzian centered in the range of 0.02 to 0.2~Hz plus a zero-centered Lorentzian for the broadband noise. The centroid frequency anticorrelates with the rms (Fig.~\ref{fig:rms_f0}), meaning that when the rms is higher (the variability amplitude is larger), the characteristic exotic variability timescale is longer.

The distinguishable feature of the new \textit{Class X} {\color{black}(Panel~f)} compared to \textit{Class V} is the uniformity of the flare timescale, which leads to a narrow peak in the averaged PSD at $0.0154\pm0.0005$~Hz{\color{black}, which does not evolve with rms}. There is also a QPO between 2 and 3~Hz, and in the averaged PSD, its centroid frequency is measured to be $2.78^{+0.07}_{-0.10}$~Hz. 
{\color{black}We note that the $\sim0.016$~Hz and $\sim2.7$~Hz features at the end of \textit{SIMS} match, within 90\% uncertainties, the frequencies of features also seen in \textit{Class X}, although there is a large difference between their rms and PSD shape.}
This is interesting because it might indicate some {\color{black}persistent and} intrinsic physical timescale in the system. We will discuss this more in Section~\ref{discussion:compare_variability_class}. 

\begin{figure}
\centering
\includegraphics[width=1.\linewidth]{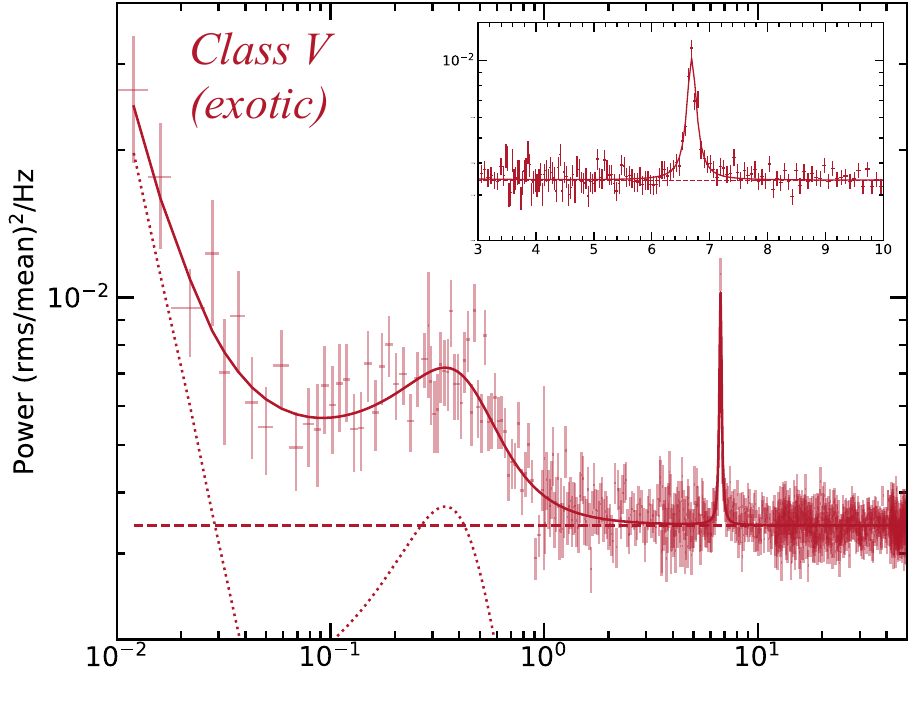}
\caption{{\color{black}The PSD that shows the highly coherent QPO. To show both the noise component at low frequencies and the highly coherent QPO, the logarithmic frequency rebinning factor is 0.2 below 3~Hz and 0.025 above 3~Hz. The QPO centroid frequency is fitted to be $6.704^{+0.013}_{-0.014}$~Hz with a Q factor of $45^{+12}_{-8}$ and a fractional rms amplitude of $4.1\pm0.2\%$ (see Wang et al., 2024b for more details).}}
\label{fig:new_qpo}
\end{figure}

\begin{figure*}
\centering
\includegraphics[width=1.0\linewidth]{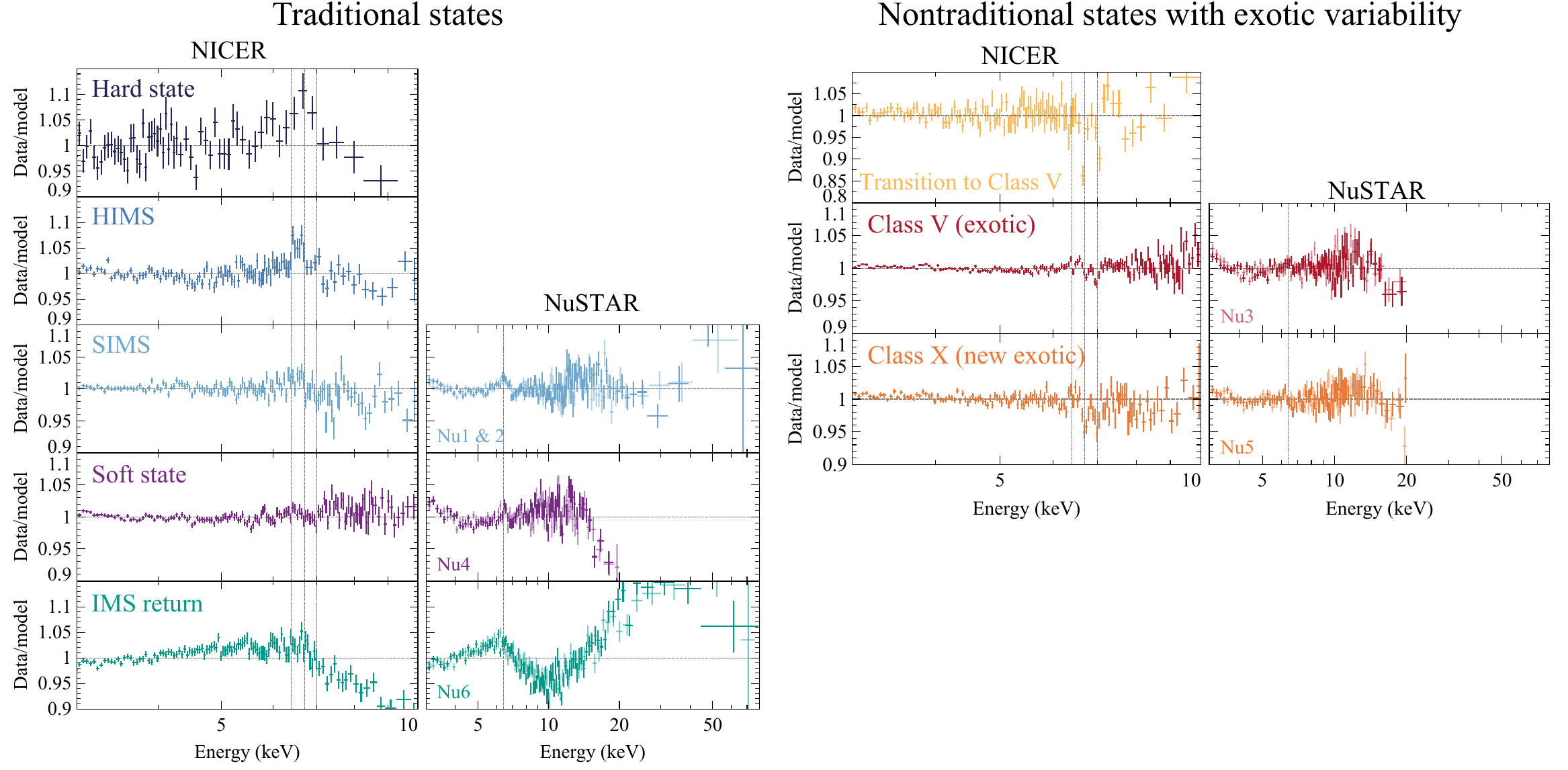}
\caption{The data-to-model ratio with the {\color{black}baseline model that includes the disk emission and the Comptonization component (see Section~\ref{sect:detailed_spec} for more details)}. (\textit{Left}) The traditional states {\color{black}with stochastic variability} including the \textit{Hard State}, \textit{HIMS}, \textit{SIMS}, and \textit{IMS Return} show iron emission line due to relativistic reflection. (\textit{Right}) The nontraditional states {\color{black}with exotic variability} including the \textit{Transition to Class V}, {\color{black}Classes V and X} that show absorption lines from highly ionized iron. The dashed lines indicate Fe K$\alpha$ at 6.4~keV, {\color{black}He}-like Fe XXV at 6.7~keV, and {\color{black}H}-like Fe XXVI at 6.97~keV in the NICER spectra, and 6.4~keV in NuSTAR spectra. {\color{black}The number after `Nu' indicates the index of the NuSTAR observation in chronological order.}}
\label{fig:ra}
\end{figure*}

In the \textit{Soft State} (Panel c), the fractional rms is very low ($\sim6\%$), and the corresponding PSD is absent of any component besides a weak power-law noise. 

{\color{black}In the dataset, we discovered a highly coherent QPO with Q factors (defined as the QPO frequency divided by the full-width-at-half-maximum) $\gtrsim50$. The QPO evolved over time with its frequency ranging between 5 and 8~Hz, appeared first on Apr 19, and disappeared on June 26 (see Fig.~\ref{fig:dpsd}). When the QPO was present, the PSD consists of the Poisson noise (PSD is flat over frequency, and is consistent with $\simeq2/\langle x\rangle$ where $\langle x\rangle$ is the averaged count rate for rms-squared normalization), red noise (PSD $\propto f^{-2}$), and an additional noise component that has either a Lorentzian centroid frequency of zero (i.e., flat-top noise) or in the range of 0.3--0.6~Hz ($>3\sigma$ away from zero; e.g., see Fig.~\ref{fig:new_qpo}). The noise component with non-zero centroid frequency appears as the low-rms extension of the Lorentzian component representing heartbeat-like exotic variability in \textit{Class V} (see Fig.~\ref{fig:rms_f0}). Therefore, for the data with a low total fractional rms $\lesssim6\%$ and a disk-dominated spectrum, we classify them as in \textit{Class V} if the centroid frequency for the noise component is non-zero, and \textit{Soft State} if the noise component centers at zero. 
A detailed analysis of the properties and evolution of the highly coherent QPO is presented in a separate paper (Wang et al., 2024b).}

In the {\color{black}\textit{IMS-return}} {\color{black}(Panel a, light blue)}, the PSD is similar to the initial \textit{HIMS} and \textit{SIMS}, with flat-top noise and a QPO in the range of 2--6~Hz (Fig.~\ref{fig:rms_f0}). 

\subsection{\color{black}{Spectral Analysis of the Iron K Band}} \label{sect:detailed_spec}

\begin{figure}
\centering
\includegraphics[width=1.\linewidth]{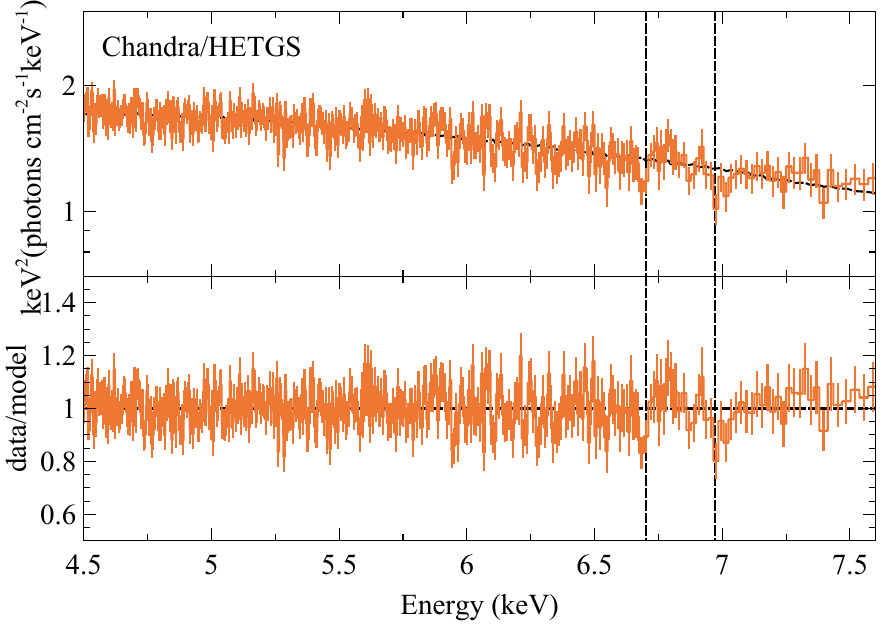}
\caption{{\color{black}The Chandra/HETG (\textit{upper}) unfolded spectrum taken in \textit{Class X}, and (\textit{lower}) the data-to-model ratio with the baseline model. The spectrum exhibits consistent absorption lines as the NICER spectrum (see Section~\ref{sect:detailed_spec} for more details).} The dashed lines indicate {\color{black}He}-like Fe XXV at 6.7~keV, and {\color{black}H}-like Fe XXVI at 6.97~keV. }
\label{fig:chandra}
\end{figure}

\begin{figure*}
\centering
\includegraphics[width=0.8\linewidth]{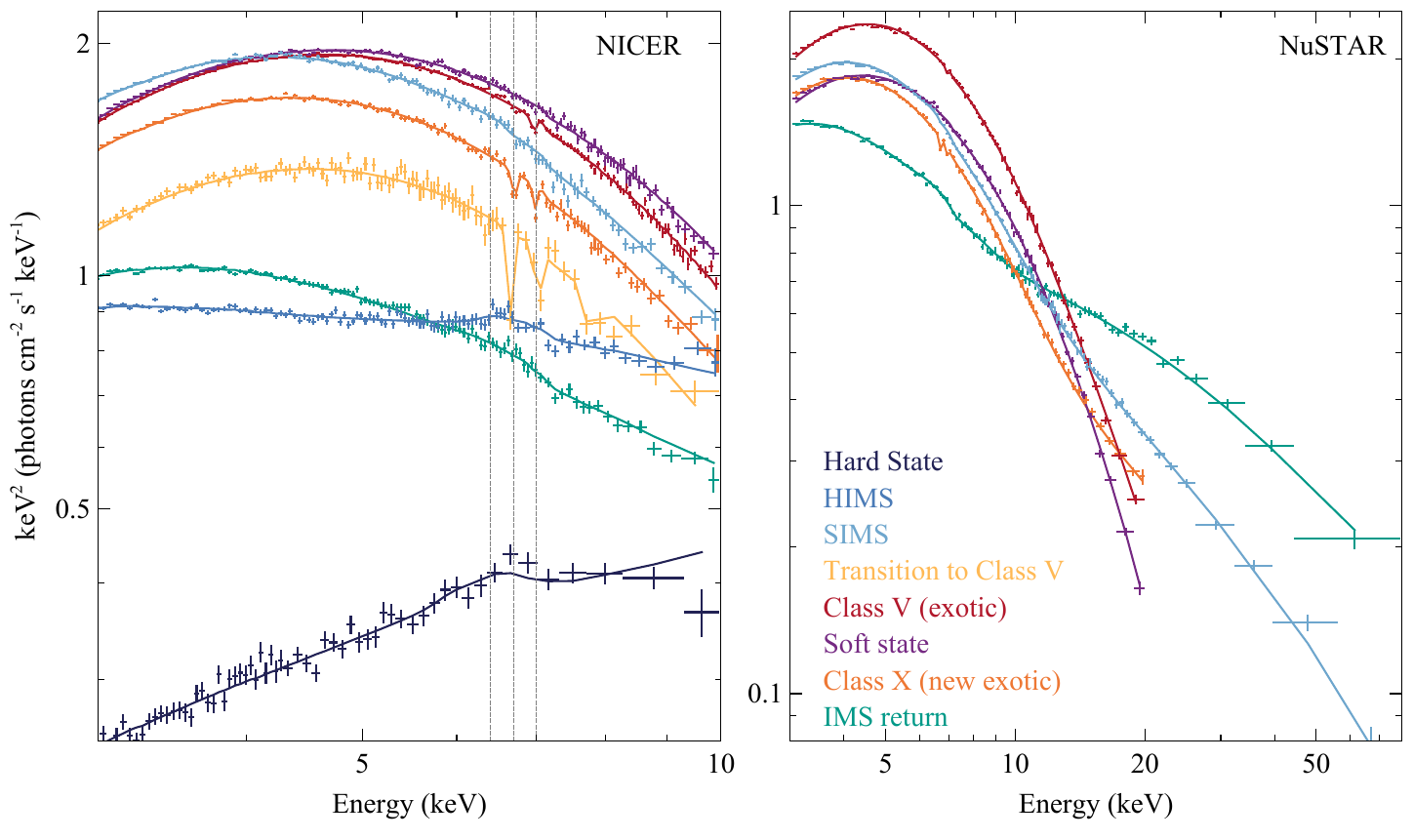}
\caption{The unfolded NICER (\textit{left}) and NuSTAR (\textit{right}) spectra using the {\color{black}final model where reflection and absorption lines are included (see Section~\ref{sect:detailed_spec} for more details).} The dashed lines indicate Fe K$\alpha$ at 6.4~keV, {\color{black}He}-like Fe XXV at 6.7~keV, and {\color{black}H}-like Fe XXVI at 6.97~keV in the NICER spectra.}
\label{fig:eeuf}
\end{figure*}

After systematically analyzing the 305 single-segment NICER flux-energy spectra (Section~\ref{sect:auto_fit}), we combine the NICER spectra in each {\color{black}of our identified states} to perform a more detailed spectral analysis, focusing especially on the iron K band. We also include NuSTAR data to cover a broad energy band and to increase the constraining power of the data. Among the 8 states, NuSTAR observations are available {\color{black}in all states besides the initial \textit{Hard State}, \textit{HIMS}, and the \textit{Transition to Class V} (this last state occurred for only one day; see Table~\ref{tab:obs}).} The NuSTAR spectra in observations 1 and 2 are combined as they are both in the \textit{SIMS}. 

First, we model the NICER and NuSTAR flux-energy spectra in all 8 states with the baseline model \texttt{TBabs*crabcorr*(diskbb+nthComp)}. The model \texttt{crabcorr} serves a NICER and NuSTAR cross-calibration purpose, multiplying each model by a power law with corrections to both the slope by $\Delta\Gamma$ and the normalization \citep{steiner2010constant}. The data-to-model ratios are shown in Fig.~\ref{fig:ra}. {\color{black}Below in this section, in order to test the significance level of the iron emission/absorption line, we add a \texttt{gaussian} line with the normalization free to be positive or negative to the baseline model. The energy, width, and normalization of the Gaussian line, and the baseline model are all free to vary.} 

{\color{black}In the states that are akin to states in typical BHXBs (i.e. \textit{Hard State}, \textit{HIMS}, \textit{SIMS}, \textit{Soft State} and \textit{IMS return}), we see a broad iron emission line in the spectrum, a canonical} signature of relativistic reflection. Another signature of reflection, the Compton hump, is clearly detected in the \textit{IMS Return} {\color{black}(when the hard Comptonized component was strongest)}. 
{\color{black}In the \textit{Soft State}, the broad iron emission line is detected at a significance level of $6\sigma$ measured with NICER and NuSTAR spectra combined.}

{\color{black}In the `exotic' states (i.e. the \textit{Class V}, \textit{Class X} and \textit{Transition to Class V})}, we detect absorption lines at energies close to the rest energies of Fe XXV (6.7~keV) and Fe XXVI (6.97~keV). {\color{black}The energies, widths, equivalent widths, and the significance levels of the significantly detected absorption lines ($>3\sigma$) are shown in Table~\ref{tab:spec_pars}. The 90\% upper limit on the blueshift is 0.08~keV, corresponding to an outflow velocity $<0.01c$.} {\color{black}We note that during the \textit{Transition to Class V}, the absorption lines at 6.7 and 6.97~keV are detected at significance levels} of 6$\sigma$ and $3\sigma$ with only 2.2~ks NICER exposure. 
The Chandra/HETG observation {\color{black}took place when the source was} in \textit{Class X}. The Chandra/HETG unfolded spectrum and the data-to-model ratio using the baseline model are shown in Fig.~\ref{fig:chandra}. We measure the Fe XXV and Fe XXVI absorption lines at $6.66^{+0.05}_{-0.04}$ and $7.00\pm0.04$~keV with both widths $<0.08$~keV. The equivalent width is $8.7^{+8.9}_{-1.2}$ and $12^{+7}_{-11}$ eV for Fe XXV and XXVI respectively. The absorption lines are therefore consistent within 90\% uncertainties with NICER. 

These results indicate that there is a broad iron emission line from relativistic reflection when there is no {\color{black}exotic} variability, and there are absorption lines from highly ionized absorbing material when there is {\color{black}exotic} variability. {\color{black}Therefore,} in our final model, we add to the baseline model (1) a relativistic reflection model \texttt{relxilllpCp}\footnote{We use \texttt{relxill} v2.2 available at \url{http://www.sternwarte.uni-erlangen.de/~dauser/research/relxill/}} \citep{garcia2014_relxill,2022MNRAS.514.3965D} for the \textit{Hard State}, \textit{HIMS}, \textit{SIMS}, \textit{Soft State}, and \textit{IMS Return}; (2) absorption lines detected above $3\sigma$ level modeled by \texttt{gaussian} for the \textit{Transition to Class V}, and \textit{Classes V and X}. The best-fitted parameter values are shown in Table~\ref{tab:spec_pars}, and the data-to-model ratios and the unfolded spectra are shown in Figs.~\ref{fig:ra_final} and \ref{fig:eeuf}, respectively. 
We note that \igr\ is a peculiar BHXB with {\color{black}sometimes} a very hot disk {\color{black}($T_{\rm in}$ can reach $\gtrsim1.5$~keV)}, so the assumptions in the reflection model {\color{black}that we use such as the low-energy break of the Comptonization component, and the geometrically thin disk assumed} are not guaranteed to hold (see Section~\ref{appendix:reflection} in the Appendix for more details). 
Therefore, the exact values of parameters constrained {\color{black}by the reflection model} 
need to be taken with caveats.

\begin{figure}
\centering
\includegraphics[width=1.\linewidth]{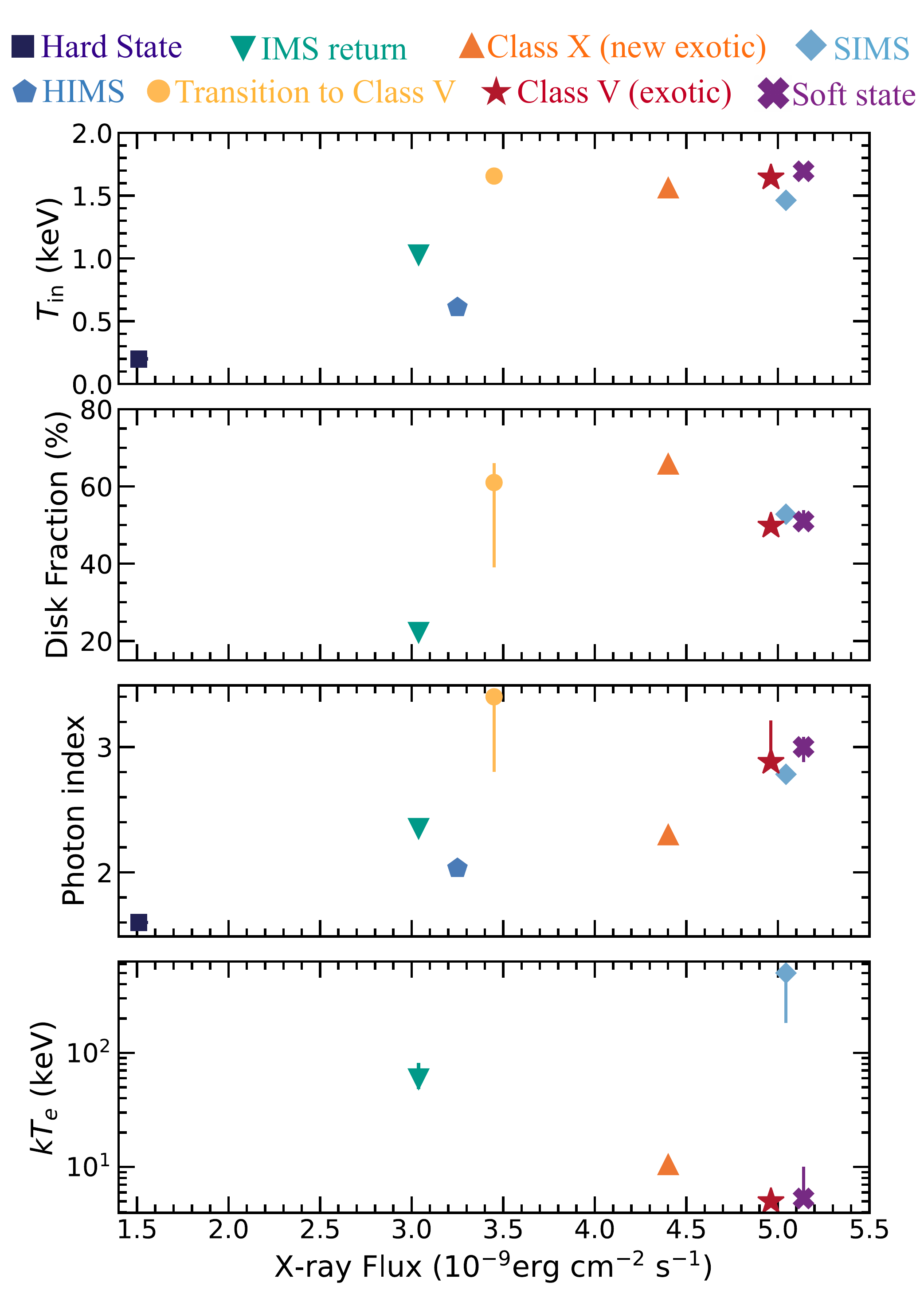}
\caption{The parameters describing the {\color{black}properties of the disk and the corona} as functions of the X-ray flux with the final model {\color{black}where reflection and absorption lines are included} (see Section~\ref{sect:detailed_spec}). The disk fraction and total X-ray flux are measured in 2--20~keV. The electron temperature $kT_{e}$ cannot be constrained and is fixed at 100~keV in \textit{Hard State}, \textit{HIMS}, and \textit{Transition to Class V}, and is therefore not shown here. }
\label{fig:pars}
\end{figure}

{\color{black}While the parameters constrained from reflection modeling (e.g. coronal height, spin, inclination, etc.) warrant caution, they do provide a good description of the reflection features, and therefore the continuum modeling is robust, especially when NuSTAR data are included. We show the parameters describing the properties of the disk blackbody and the corona} in each state at different X-ray flux in Fig.~\ref{fig:pars}. As found in the single-segment NICER spectral fits (Fig.~\ref{fig:mjd}), the disk temperature is $\lesssim1$~keV in the \textit{Hard State}, \textit{HIMS}, and \textit{IMS Return}, and varies in the range of 1.5 to 2~keV in {\color{black}the other 5 states, when the disk fraction is $>50\%$. These 5 states also have a soft coronal spectrum with a high photon index $\gtrsim2$, {\color{black}which further solidifies the identification of the \textit{SIMS} and \textit{Soft State} if mapped to traditional spectral states of BHXBs.} We note that the electron temperature of the corona is very low in \textit{Class V}, \textit{Class X}, and the \textit{Soft State}, around 5--10~keV. This has been found also in \grs\ (e.g., \citealp{2011ApJ...737...69N}).}

{\color{black}The global parameters that should not change in time are tied between epochs, and as such,} they are constrained with more confidence than the parameters from reflection spectroscopy from individual states. The column density of galactic absorption is constrained to be {\color{black}$N_{\rm H}=(1.537\pm0.002)\times 10^{22}$cm$^{-2}$}, consistent with previous measurements (e.g., \citealp{2017ApJ...851..103X,2018MNRAS.478.4837W}). The fitted inclination angle via reflection spectroscopy is {\color{black}$\left(24\pm4\right)^\circ$}. Previous measurements from reflection spectroscopy using NuSTAR data in the hard state resulted in $i=(37^{+3}_{-4})^\circ$ and $\left(45.3\pm0.7\right)^\circ$ \citep{2017ApJ...851..103X,2018MNRAS.478.4837W}, also suggest a low inclination for the inner disk producing the reflection.

\subsection{{\color{black}A Summary of the Key Properties of Each Accretion State}} \label{sect:states}

{\color{black}In Section~\ref{sect:tools} we describe the broadband continuum shape (i.e. whether corona- or thermal-dominated) and the shape of the light curves (i.e. whether showing exotic variability) in order to classify the complex phenomenology of \igr\ into eight states. Some of those states are akin to states in traditional BHXBs (namely, the \textit{Hard State}, \textit{HIMS}, \textit{SIMS}, \textit{Soft State} and \textit{IMS return}), and then when the source is thermal disk-dominated, it can sometimes take excursions into states with very complex and exotic variability (namely, the \textit{Transition to Class V}, \textit{Class V} and \textit{Class X}). Here we summarize the key properties of each of these states, and particularly describe the emission/absorption line structure and the PSD structure of each state. The states listed below are in the order in which they appeared for the first time in this outburst (see Table~\ref{tab:obs}). }{\color{black}The quoted measured quantities using flux-energy spectra can be found in Table~\ref{tab:spec_pars}.}


\begin{itemize}
    \item \textit{Hard State}: {\color{black}As in typical BHXBs, the variability in this state} is stochastic with a high averaged fractional rms of 27\%, and the PSD consists of a Type-C QPO at $\sim0.3$~Hz on top of flat-top noise. The flux-energy spectrum contains a cool disk with disk temperature {\color{black}$T_{\rm in}=0.20\pm0.02$~keV}, a corona with a hard spectrum (photon index {\color{black}$\Gamma=1.60\pm0.02$}), and a broad iron {\color{black}emission} line due to relativistic reflection.
    \item \textit{Hard-Intermediate State} (\textit{HIMS}): The variability is still stochastic while the fractional rms has decreased to 12\%. Both the frequencies of the Type-C QPO and the low-frequency break of the flat-top noise increase compared to the \textit{Hard State}. The QPO frequency is between 2.5 and 5~Hz. The disk temperature increases to {\color{black}$0.61^{+0.07}_{-0.02}$~keV}, and the coronal spectrum is softer than it was, with {\color{black}$\Gamma=2.033\pm0.013$}. The broad iron {\color{black}emission} line is present from reflection.
    \item \textit{Soft-Intermediate State} (\textit{SIMS}): The fractional rms further declines to 8\%, and a Type-B QPO at 2--3~Hz is present. {\color{black}At the end of \textit{SIMS} (Mar 26),} the lightcurve started to show flares and modulations {\color{black}that are signs of emerging exotic variability } on a timescale of $62^{+4}_{-8}$~s, corresponding to a narrow peak at $0.016^{+0.002}_{-0.001}$~Hz in the PSD. The disk becomes hotter {\color{black}than it was} with a disk temperature of {\color{black}$1.463^{+0.005}_{-0.004}$~keV}, and the {\color{black}coronal spectrum} further softens {\color{black}to} {\color{black}$\Gamma=2.782^{+0.012}_{-0.015}$}. The broad iron {\color{black}emission} line is detected in both NICER and NuSTAR spectra.
    \item \textit{Transition to Class V}: The lightcurve shows that \textit{Class V} exotic variability developed during 4~hours in this transitional state, {\color{black}while} the flux decreased to {\color{black}half that of} the \textit{SIMS} peak (see Fig.~\ref{fig:hid_cc}a). Due to this exotic variability, the fractional rms increases to 16\%, and the power increases generally across frequencies from 0.002 to $\sim10$~Hz. The disk is still very hot with {\color{black}$T_{\rm in}=1.656^{+0.010}_{-0.020}$~keV}, and the coronal spectrum is soft with {\color{black}$\Gamma>2.8$}. Strong absorption lines close to the rest energies of Fe XXV and Fe XXVI are detected at $6\sigma$ and $3\sigma$ confidence levels. 
    \item \textit{Class V}: Exotic {\color{black}flaring} variability is evident in the lightcurves. The PSD contains a broad component with centroid frequency in the range of 0.02 to {\color{black}0.5~Hz}, resulting from the irregularity of the exotic variability. The irregular variability pattern and the broad component in the PSD are similar to \textit{Class V} in \citet{2017MNRAS.468.4748C}, corresponding to class $\mu$ in \grs\ \citep{2000A&A...355..271B}. We will discuss more on this in Section~\ref{discussion:compare_variability_class}. The averaged fractional rms is {\color{black}12\%}. The disk temperature is {\color{black}$1.644^{+0.005}_{-0.003}$~keV} with a high disk fraction of {\color{black}$49.8^{+0.3}_{-0.7}$\%}, and the photon index is {\color{black}$2.88^{+0.33}_{-0.02}$}. An iron XXVI absorption line is detected at a $5\sigma$ confidence. 
    \item \textit{Soft State}: No exotic variability can be seen in the lightcurves, and the averaged fractional rms is only {\color{black}5\%}, the lowest among the states. Starting from Apr 19, {\color{black}a highly coherent QPO appeared (Wang et al., 2024b)}. The disk temperature is {\color{black}$1.694^{+0.013}_{-0.016}$~keV}, and the disk fraction is {\color{black}$50.9^{+2.9}_{-0.4}$\%}. The {\color{black}coronal spectrum} is soft with {\color{black}$\Gamma=3.00^{+0.08}_{-0.12}$}, and an iron emission line is detected at $6\sigma$ significance. The low rms, hot disk domination, and a {\color{black}soft coronal spectrum} are the features that are in agreement with a traditional soft state (see also the discussion in Section~\ref{discussion:compare_variability_class}). 
    \item \textit{Class X}: {\color{black}Exotic, large-amplitude, near-sinusoidal variability} is prominent in the lightcurves, and the fractional rms increases to 24\%. The variability amplitude can change within 500~seconds (see Fig.~\ref{fig:lc}g--h), but the uniformity of the flare timescale leads to a narrow peak in the averaged PSD at $0.0154\pm0.0005$~Hz. This class {\color{black}has never been seen before in either this source or \grs}, and we will discuss it more in Section~\ref{discussion:compare_variability_class}. With a disk temperature of {\color{black}$1.562^{+0.003}_{-0.004}$~keV}, the disk fraction is {\color{black}$65.8^{+0.7}_{-0.6}$\%}, the highest among all the states. Both Fe XXV and XXVI absorption lines are detected in NICER and Chandra/HETG spectra.
    \item \textit{Intermediate State Return} (\textit{IMS Return}): The variability is stochastic, and the PSD shape is similar to the initial \textit{HIMS}. This state connects the initial \textit{HIMS} and \textit{SIMS} in the mimicked HID and color-color diagram (Fig.~\ref{fig:hid_cc}). Compared to previous states, the disk temperature has dropped to {\color{black}$1.028^{+0.015}_{-0.008}$~keV}, along with a lower disk fraction of {\color{black}$22.2^{+0.9}_{-0.6}$\%}. The {\color{black}coronal spectrum} is still soft with {\color{black}$\Gamma=2.35^{+0.03}_{-0.04}$}. Both reflection signatures, the broad iron line, and the Compton hump are prominent in the NICER and NuSTAR spectra.
\end{itemize}


\section{Discussion} \label{discussion}

\subsection{Comparison of Variability Classes} \label{discussion:compare_variability_class}

{\color{black}Previously, nine variability classes were defined for \igr\ in its 2011--2013 outburst \citep{altamirano2011faint, 2017MNRAS.468.4748C}. In this section, we will compare the variability classes we identify for this 2022 outburst with those in previous works.}

{\color{black}In the 2022 outburst, we observed \textit{Class V} identified based on the repeated, sharp, high amplitude flares in the lightcurves and the PSD shape.} We note that a QPO at $\sim4$~Hz with a Q factor of $\sim3$ was observed in the \textit{Class V} in the 2011 outburst (Fig.~11 in \citealp{2017MNRAS.468.4748C}). 
{\color{black}While \textit{Class V} variability was observed previously, here we see for the first time that the timescale of variability can evolve dramatically, and scales with the fractional rms of the source (Fig.~\ref{fig:rms_f0}). }

We define a new \textit{Class X} because of several properties different from {\color{black}the} previously identified nine classes. Although the uniformity of the exotic variability timescale is reminiscent of the heartbeat class (\textit{Class IV} in \igr\ and \textit{Class $\rho$} in \grs), the flare patterns are more symmetric than the typical heartbeat (slow rise and quick decay). In the PSD, there is sometimes a QPO at 2--3~Hz, while in \textit{Class IV}, previously no QPO was detected \citep{2017MNRAS.468.4748C}, or the QPO was at 6--10~Hz \citep{altamirano2011faint}. We also find that in \textit{Class X}, the exotic variability amplitude can vary within even 500~seconds while maintaining the variability timescale (see Fig.~\ref{fig:lc}g and h). When the variability amplitude is relatively small, the variability pattern is {\color{black}not only similar to \textit{Class III} (\textit{Class $\nu$} in \grs), but also similar to the end of SIMS when the source first started to show flares and modulations. The modulation produces a peak in the PSD at $\sim0.016$~Hz, in addition to a Type-B QPO at $\sim2.7$~Hz. The two characteristic frequencies are respectively consistent within 90\% uncertainties at the end of SIMS and in Class X (see Section~\ref{sect:psd} and Fig.~\ref{fig:rms_f0}). This suggests that Class X can be regarded as a high-rms extension of the structured variability that starts at the end of SIMS. }


{\color{black}We identify the \textit{Soft State} based upon the lack of exotic variability, the lowest fractional rms of 6\%, the hot disk domination, and a very soft coronal spectrum. There are two classes identified in \citet{2017MNRAS.468.4748C} that also show no exotic variability in the lightcurves -- \textit{Class I} and \textit{Class II}.
The PSD in \textit{Class I} is similar to the intermediate state here, with a broadband noise at 1--10~Hz, and a QPO at $\sim5$~Hz. 
We notice the PSD in our \textit{Soft State} in the NICER hard band 4.8--9.6~keV is similar to the one in \textit{Class II} using RXTE data (2--60~keV), both lack any power above $\sim1$~Hz. However, in our \textit{Soft State}, the PSD sometimes shows a highly coherent QPO at 5--8~Hz
(Wang et al., 2024b), which was not detected in \textit{Class II}.} 

\subsection{\igr\ as the Bridge Between \grs\ and Normal BHXBs} \label{discussion:bridge}

In this work, we find that in the 2022 outburst, \igr\ went through a traditional hard state and intermediate state, and then entered an exotic soft state where it sometimes exhibits heartbeat-like variability in the light curves. This transition from traditional BHXB states to states showing exotic variability has also been suggested in its previous two outbursts in 2011 and 2016 \citep{2014ApJ...783..141P,2018MNRAS.478.4837W}.

In a seminal work, {\color{black}for \grs,} \citet{2000A&A...355..271B} identified 12 variability classes and three basic states and suggested that the variability classes are produced by transitions between the three basic states with certain patterns. Although they found some similarities between the three basic states and canonical BHXB accretion states including the soft state and intermediate state, in the variability class \grs\ spends most of its time in (Class $\chi$), the spectrum is the hardest and the lightcurve shows stochastic variability, it is still not as hard as the canonical hard state of BHXBs. Therefore, the accretion state landscape in \grs\ is significantly more challenging to understand due to the more complex phenomenology. 

In summary, \igr\ shares some variability classes with \grs\ that are not seen in other BHXBs, while having an outburst recurrence rate {\color{black}(see a summary in Section~\ref{intro})} and evolution pattern in outbursts similar to normal BHXBs. It can therefore be regarded as a bridge between the most peculiar BHXB \grs\ and normal BHXBs. 

\begin{figure*}
\centering
\includegraphics[width=1.\linewidth]{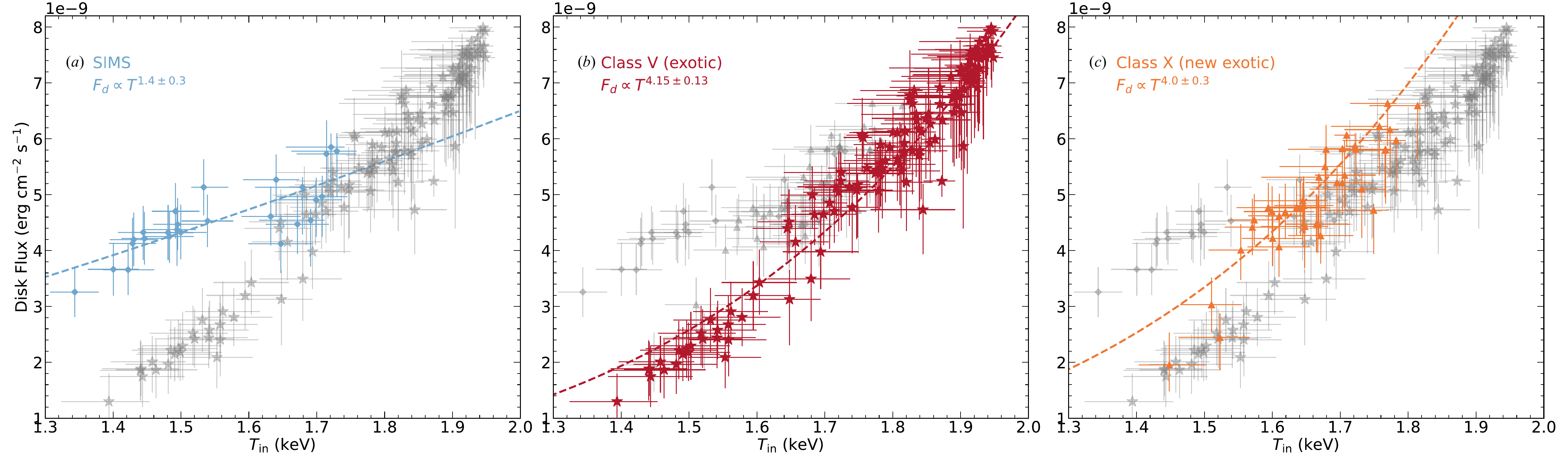}
\caption{{\color{black}The disk flux (0.1--200~keV) versus disk temperature $T_{\rm in}$ in the segment-based fits of the broadband continuum (Section~\ref{sect:auto_fit}) in three different states. The dashed lines are the best-fits with the model $F_d\propto T^n$.}}
\label{fig:F-T}
\end{figure*}

\subsection{The Nature of Heartbeats} \label{discussion:heartbeat nature}

\subsubsection{A brief review of radiation pressure instability}\label{discussion:radiation_pressure_instability}

{\color{black}Although it is generally accepted that the heartbeat-like variability is due to some disk instability, the nature of that instability is unclear. In the standard Shakura \& Sunyaev disk, there are two types of instability that might lead to limit-cycle oscillations in the lightcurves: hydrogen ionization instability and radiation pressure instability (e.g., \citealp{done2007modelling,2011MNRAS.414.2186J}). The ionization instability takes place in the outer disk and may explain the outburst and quiescence behavior of BHXBs, i.e., lightcurve variability on a long timescale (hundreds of days). On the other hand, radiation pressure instability sets in at a relatively higher mass accretion rate, and thus the inner disk becomes radiation pressure dominated. This could then lead to thermal-viscous limit cycles, and therefore is a widely accepted physical mechanism for driving the heartbeat-like variability \citep{1997ApJ...488L.109B,2000ApJ...542L..33J,2000ApJ...535..798N,2011ApJ...737...69N}. }

{\color{black}One way to understand the radiation pressure instability is through the `S-curve' when plotting the mass accretion rate $\dot{m}$ versus the disk surface density $\Sigma$ at a certain disk radius (e.g., \citealp{done2007modelling}). From bottom to top, the S-curve consists of 3 branches: a stable branch when heating is proportional to the gas pressure balanced by radiative cooling, an unstable branch when radiation pressure dominates (`Lightman-Eardley instability'; \citealp{1974ApJ...187L...1L}), and another stable branch when advective cooling is effective to balance the heating (`slim disk' solution; \citealp{1988ApJ...332..646A}). That is to say, when the inner disk is radiation pressure dominated and $\dot{m}$ is on the middle unstable branch, a limit-cycle instability is expected. Over each limit cycle, the mass accretion rate oscillates at each disk radius (switching between the two stable branches), resulting in a `density wave'.
Observational pieces of evidence supporting this hypothesis include the correlated limit-cycle timescale and the disk inner radius \citep{1997ApJ...488L.109B}, and extensive phase-resolved spectroscopy analysis of the heartbeats (e.g., \citealp{2011ApJ...737...69N,2016ApJ...833..165Z}). }

{\color{black}We observe a persistent heartbeat timescale of $\sim60$~seconds at the end of \textit{SIMS} and in \textit{Class X} (see Section~\ref{sect:psd} and Fig.~\ref{fig:rms_f0}). A back-of-envelope estimate of the limit-cycle duration is therefore the viscous timescale for the accretion disk} \citep{1997ApJ...488L.109B}, which is given by

\begin{equation}
	t_{\rm visc}=60\, {\rm s} \left(\frac{M}{10M_\odot}\right)\left(\frac{R/R_{\rm g}}{46}\right)^{3/2}\left(\frac{\alpha}{0.03}\right)^{-1}\left(\frac{H/R}{0.1}\right)^{-2},
	\nonumber
\end{equation}
where $M$ is the mass of the black hole, $R$ is the radius in the disk, $R_{\rm g}$ is the gravitational radius $GM/c^2$, $\alpha$ is the viscosity parameter, and $H$ is the vertical scale height of the disk (e.g., \citealp{2002apa..book.....F}). {\color{black}A viscous timescale of 60~seconds can be explained, e.g., by the parameter combination including a black hole mass of $10M_\odot$, $\alpha=0.03$, $H/R=0.1$, and $R=46R_{\rm g}$; or, if the black hole mass is $3M_\odot$, $R=102R_{\rm g}$. } 

\subsubsection{Did \igr\ reach the Eddington limit?}\label{discussion:Eddington}

{\color{black}The Eddington ratio threshold for radiation pressure instability to occur is very uncertain (could range from 6\% $L_{\rm Edd}$ to near Eddington), fundamentally because the S-curve depends on a variety of physical assumptions and conditions (e.g., \citealp{1991PASJ...43..147H,2000ApJ...542L..33J,2002ApJ...576..908J}). It is therefore natural to find an observational threshold of the Eddington ratio where heartbeats are seen to constrain the underlying accretion disk physics.}

{\color{black}So far, heartbeat-like (repetitive, high amplitude, and structured) variability has been observed in both accreting black hole and neutron star systems and can be a multiwavelength phenomenon. (1) Two BHXBs including \grs\ and \igr. In \grs, besides X-ray, heartbeat-like variability was also seen in radio and infrared (e.g., \citealp{1997MNRAS.292..925P,1997MNRAS.290L..65F}). As the mass and distance of \grs\ are known, it reaches 80\%--90\% of the Eddington luminosity \citep{2011ApJ...737...69N}. (2) Three neutron star low-mass X-ray binaries including GRO~J1744--28 (known as the `Bursting Pulsar'; \citealp{1996Natur.379..799K,2014ApJ...796L...9D}), MXB~1730--335 (the `Rapid Burster'; \citealp{2015MNRAS.450L..52B}) and Swift~J1858.6--0814 \citep{2023arXiv230300020V}. The heartbeat-like (\grs-like) variability in neutron stars is also sometimes referred to as Type-II X-ray bursts. The luminosity reached near Eddington, $\lesssim20$\% and 40\%~$L_{\rm Edd}$ for the three sources respectively. We note that the criterion for radiation pressure instability to occur in neutron star systems is when the magnetospheric radius is larger than the neutron star radius \citep{2019A&A...626A.106M}. This means that the Eddington ratio is not the only key parameter, but also the magnetic field strength. It was shown that all three neutron star X-ray binaries satisfy this condition for heartbeat-like variability to take place \citep{2023arXiv230300020V}. (3) An ultraluminous X-ray source in NGC~3621 with an unclear compact object nature \citep{2020ApJ...898..174M}. }


{\color{black}Then the challenging part of finding an observational Eddington ratio threshold arises -- we only have two BHXBs exhibiting heartbeats, and for \igr, we do not know the black hole mass (from the mass function) or distance (e.g., from parallax) to estimate the Eddington ratio. The position of \igr\ on the radio versus X-ray luminosity diagram suggests a distance between $\sim$11 and $\sim$17~kpc for it to lie on the track followed by other BHXBs \citep{2011A&A...533L...4R}. However, it is also possible for \igr\ to not follow the typical relation, as was found for \grs. A black hole mass in the range of 8.7 to 15.6~$M_\odot$ is obtained under the two-component advective flow framework \citep{2015ApJ...807..108I}. On the other hand, there are several BHXBs reaching close to the Eddington limit and do not exhibit heartbeat-like variability (e.g., XTE~J1550--564, 4U~1543--47; \citealp{2003ApJ...595.1032R,2021ATel14708....1N}). Therefore, before heartbeats in \igr\ were discovered, it was proposed that \grs\ was the unique BHXB that showed heartbeats because it was the only one that had spent any considerable time \textit{above} the Eddington limit \citep{done2004grs}. Could this also be the case for \igr? }

{\color{black}In Fig.~\ref{fig:F-T}, we show the disk flux $F_{d}$ versus the disk temperature $T_{\rm in}$ in the segment-based fits of the broadband continuum (see Section~\ref{sect:auto_fit}) in three different states. We also perform fits with the model $F_d\propto T^n$ (i.e., the disk luminosity $L\propto T^n$), where the power-law index $n$ could indicate the nature of the accretion flow. The standard Shakura \& Sunyaev disk predicts $L\propto T^4$, and the advection-dominated accretion flows including the `slim disk' follows $L\propto T^2$ \citep{2000PASJ...52..133W}. 
We find that in both variability classes showing heartbeat-like variability, the disk luminosity-temperature (L-T) relation is consistent with that of a thin disk ($n=3.7\pm0.2$ and $4.0\pm0.3$ in Class V and X). The indication that heartbeat-like variability is associated with a thin disk is consistent with previous conclusions from a theoretical point of view \citep{2000ApJ...535..798N}. We also notice previous phase-resolved spectroscopy of \grs\ found that the disk forms a loop in the L-T diagram over heartbeat cycles \citep{2011ApJ...737...69N}. On the other hand, in the SIMS, the disk has a flatter L-T relation, with $n=1.4\pm0.3$, more in line with a slim disk that is expected at a relatively high Eddington ratio ($\gtrsim30\%$; \citealp{2010A&A...521A..15A}). It is also consistent with $n=4/3$ when the mass accretion rate is constant and the disk inner radius is variable, expected when the local Eddington limit is reached \citep{2009ApJ...696.1257L}. In either interpretation of the flatter L-T relation, a relatively high mass accretion rate ($\gtrsim30\%$ of the Eddington rate) is suggested. }

\begin{figure}
\centering
\includegraphics[width=1.\linewidth]{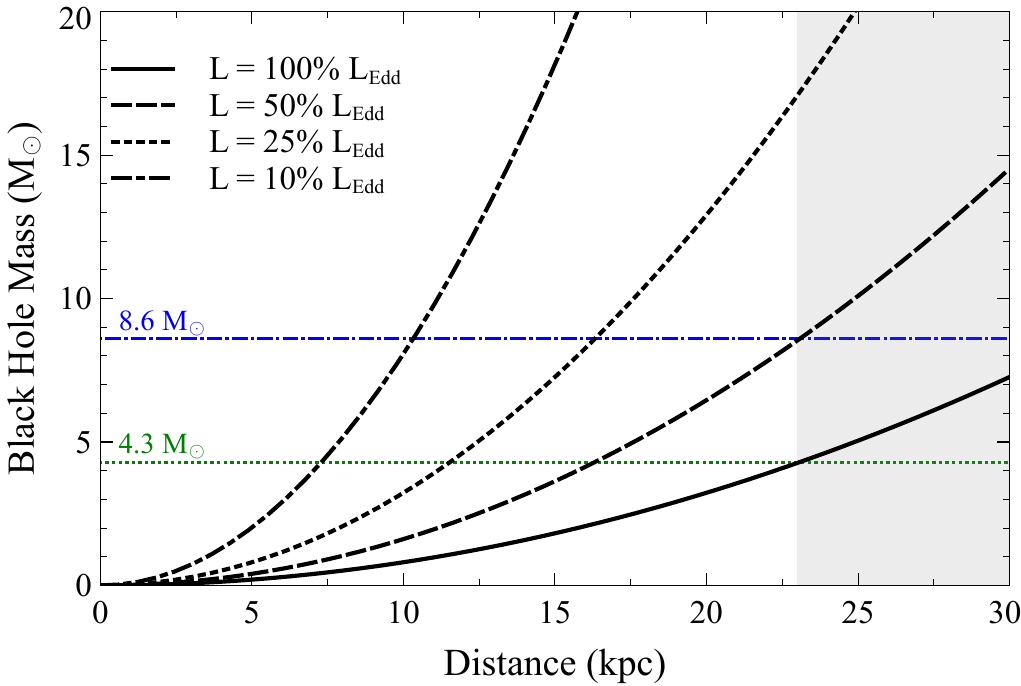}
\caption{{\color{black}The black hole mass and distance if \igr\ accretes at different Eddington ratios at the SIMS peak when the bolometric flux is $\sim8.5\times10^{-9}$ erg cm$^{-2}$ s$^{-1}$. The gray region indicates the distance at which \igr\ would be outside of the `edge' of the stellar disk in our galaxy (see Section~\ref{discussion:Eddington} for details).
}}
\label{fig:M-d}
\end{figure}

{\color{black}In our observing campaign, we caught the precursor of the heartbeats: the end of \textit{SIMS} and \textit{Transition to Class V}, during which the flux dropped gradually from the SIMS peak to half of that. Considering the suggested slim disk nature in the SIMS and the idea of the `S-curve', we have a plausible explanation for this precursor behavior. In SIMS, the inner disk lies on the stable upper branch in the S-curve which corresponds to the slim disk solution. Then, the mass accretion rate decreased, the inner disk thus entered the middle unstable branch, and heartbeat-like variability (but the variability amplitude is relatively small at the end of SIMS, and is irregular in the \textit{Transition to Class V}) started to show. Then, in \textit{Class V} and \textit{Class X}, the inner disk started to exhibit limit-cycle variabilities, switching between the upper and lower stable branches. The heartbeat-like variability is more structured and regular compared to the precursor phase. The flux in these two classes is also in the middle of the SIMS peak and the \textit{Transition to Class V}.}

{\color{black}The flux peak in SIMS is near the flux peak in the entire outburst. The measured unabsorbed flux (in units of $10^{-9}$ erg cm$^{-2}$ s$^{-1}$) at the SIMS peak is $\sim8.5$ (0.1--200~keV; flux shown in Fig.~\ref{fig:hid_cc}a is in 1--10~keV). Assuming that \igr\ at the SIMS peak emits at different Eddington ratios, the black hole mass and distance are shown in Fig.~\ref{fig:M-d}. Based on the coordinates of \igr\ (galactic longitude $l=349.52^\circ$, and latitude $b=2.21^\circ$), assuming the distance to the galactic center is 8~kpc, and the radius of the `edge' of the stellar disk in our galaxy is 10--15~kpc \citep{2016ARA&A..54..529B}, the upper limit of distance to \igr\ if it is within the stellar disk is $\sim23$~kpc. Then, in the case of emitting at 100\% Eddington, the black hole mass needs to be $<4.3$~$M_\odot$. This would make \igr\ one of the least massive black holes known. This limit could be brought up to $<8.6$~$M_\odot$ when adopting a bolometric correction factor of 2.}

{\color{black}Another empirical benchmark of the Eddington ratio is that the luminosity during the soft-to-hard state transition has a mean value of 2--3\% $L_{\rm Edd}$ \citep{2010MNRAS.403...61D,2003A&A...409..697M,tetarenko2016watchdog}. As the SIMS peak has $\sim$2 times the flux of the soft-to-hard transition, assuming a slim disk in the SIMS expected $\gtrsim30\%$~$L_{\rm Edd}$, the soft-to-hard transition in this outburst was at $\gtrsim15\%$~$L_{\rm Edd}$, which is $\sim1.8\sigma$ above the mean value using the standard deviation in \citet{tetarenko2016watchdog}.}

{\color{black}In summary, our data suggest that \igr\ in its 2022 outburst reached a relatively high Eddington ratio because of the L-T relation in SIMS. Depending on the accretion disk model, this can be, e.g.,  $\gtrsim30\%$~$L_{\rm Edd}$ \citep{2010A&A...521A..15A}, but depends on several parameters such as the magnetization (e.g., \citealp{2022A&A...659A.194M}). We cannot rule out the possibility that \igr\ reached a super-Eddington luminosity based on available black hole mass and distance constraints. Revisiting the conclusion in \citet{done2004grs}, it is therefore still possible that a super-Eddington luminosity is a necessary condition for heartbeat-like variability to be present. If super-Eddington accretion is a necessary condition, we see new evidence that it alone is not a sufficient condition. In the mimicked HID (Fig.~\ref{fig:hid_cc}a), the \textit{Soft State} reaches a similar X-ray flux as the two exotic variability classes (\textit{Classes V and X}). This might suggest that it is not just the mass accretion rate that determines whether exotic variability occurs, but rather another factor is required. This other factor could perhaps be the timescale for the inner accretion disk to respond to mass accretion rate changes in the outer disk. 
We also cannot rule out the possibility that in the \textit{Soft State}, heartbeat variability is present at other wavelengths rather than the X-ray. 
For example, in Swift~J1858.6--0814, the neutron star system where heartbeat-like variability was discovered recently, the heartbeat is the most manifest in the infrared band and is much less discernible in X-ray. On the other hand, if \igr\ reached only sub-Eddington, it is also a mystery what factor sets \igr\ apart from other BHXBs to exhibit heartbeat-like variabilities. }

\subsubsection{Beyond the radiation pressure instability model}\label{discussion:beyond}

If {\color{black}the heartbeat-like variability is} not related to a high Eddington accretion ratio, the known large disk size and the longest orbital period in \grs\ (33.5~days; \citealp{reid2014grs1915}) might be the distinguishing factor. {\color{black}In addition, the Bursting Pulsar has a very long orbital period of 11.8~days \citep{1996Natur.381..291F}. 
For \igr,} one {\color{black}piece of} supporting evidence comes from an empirical study of the relationship between the quiescent luminosity and the orbital period \citep{2011ApJ...734L..17R,2012MNRAS.422L..91W}. {\color{black}Assuming this relationship holds,} \igr\ is inferred to also have a long orbital period of $>4$~days for a distance of 10~kpc (can even be tens of days if the distance is actually larger). {\color{black}However, there are several BHXBs with orbital periods longer than 4~days (see Table~13 in \citealp{tetarenko2016watchdog}). {\color{black}If \igr\ indeed has the second longest orbital period, it has to be $\gtrsim480$~hours~$=20$~days. Then, if we assume the linear relationship between quiescent luminosity and orbital period holds (the original empirical relationship extends to an orbital period $\sim100$~hours; \citealp{2011ApJ...734L..17R}), the distance to \igr\ inferred is $\gtrsim22$~kpc, which means \igr\ is at the closest on the edge of the stellar disk in our galaxy. In addition, we notice} the mechanism that could give rise to heartbeat-like variability in a large accretion disk is not clear. Theory predicts that a longer orbital period corresponds to a larger peak mass accretion rate \citep{2002ApJ...565.1107P}, and a systematic study of low-mass X-ray binaries has revealed such correlation \citep{2010ApJ...718..620W}, which relates back to the high Eddington ratio scenario. It was also proposed that the accretion disk in a system with a long orbital period suffers from instabilities in the disk's vertical structure \citep{1999MNRAS.309..561Z,2016Natur.529...54K}.}

Another plausible hypothesis {\color{black}for the heartbeat 
is \textit{disk tearing}. In this scenario, if there is initially a tilted accretion disk (misaligned black hole spin axis and the binary orbital axis), Lense-Thirring precession warps the disk, and the disk breaks into discrete rings each precessing at a different rate. In the hydrodynamical simulation from \citet{2021ApJ...909...81R,2021ApJ...909...82R}, disk tearing leads to mass accretion rate variations that follow a heartbeat pattern.} {\color{black}One finding that might support a disk tearing scenario is that the measured inclination angle from reflection off the inner disk suggests a lower inclination ($\sim30$--40 degrees; see Section~\ref{sect:detailed_spec}), while the disk winds in the soft state of BHXBs 
are usually expected to be confined to the equatorial plane and therefore in nearly edge-on systems (60--80 degrees; \citealp{2012MNRAS.422L..11P}).
If the low inclination from disk reflection indicates the inner disk inclination and therefore the black hole spin axis, and the high inclination from the high velocity outflow indicates the outer disk inclination and therefore the orbital axis, this discrepancy might suggest a tilted disk in the system. In \grs, \citet{2016ApJ...833..165Z} performed spectral modeling over heartbeat cycles, and found through reflection spectroscopy that the inner disk inclination varied by $\sim10^\circ$ over a heartbeat cycle, but there is no evidence for misalignment between the inner and outer disk. }

\subsubsection{Interplay between heartbeats and iron emission/absorption line}\label{discussion:emission_absorption}

{\color{black}One important finding from this NICER and NuSTAR campaign is the interplay between iron emission and absorption lines during this outburst. More specifically, we observe (1) that in the states that are akin to typical BHXBs states (i.e. \textit{Hard State}, \textit{HIMS}, \textit{SIMS}, \textit{Soft State} and \textit{IMS return}), we see a broad iron emission line in the spectrum, a canonical signature of relativistic reflection; and (2) in the `exotic' states (i.e. the \textit{Class V}, \textit{Class X} and \textit{Transition to Class V}), we detect absorption lines at energies close to the rest energies of Fe XXV (6.7~keV) and Fe XXVI (6.97~keV). In this section, we discuss the implication of this interplay between exotic variability and iron emission/absorption line.}

{\color{black}First, we will discuss the wind-driving mechanism. In general, disk winds in BHXBs could be driven via thermal, radiative, and/or magnetic pressure (see a review in \citealp{2016AN....337..512P} and references therein). 
In the high-Eddington-accretion scenario, high velocity radiation-driven outflows are expected from the innermost regions. In the disk tearing scenario (see Section~\ref{discussion:beyond}), a magnetically-driven wind is seen in general relativistic magnetohydrodynamic (GRMHD) simulations of a tilted disk with a high velocity of 0.02--0.2~$c$ launched between 10 and 500~$R_{\rm g}$ \citep{liska2019bardeen}. In our case, the blueshifts of the significantly detected absorption lines are at most 0.08~keV, indicating a very slow outflow velocity ($<0.01c$). Therefore, if the wind in this dataset is radiatively driven or magnetically driven as in \citet{liska2019bardeen}, the high-velocity component is likely too hot and over-ionized. Alternatively, the wind may be thermally driven, but the exact relationship between thermal winds and heartbeat-like variabilities is not clear.
Thermal winds generally require the inner disk to be able to geometrically illuminate the outer disk, the luminosity to be high enough, and the spectrum to be hard enough that gas in the outer disk can be heated to exceed the local escape speed \citep{2010ApJ...715.1191R}. If we assume the outflow velocity equals the escape velocity at the wind launching radius, the outflow is launched at a radius larger than $2\times10^4$~$R_{\rm g}$. }



{\color{black}Then it becomes puzzling why we do not observe significant absorption lines in \textit{SIMS} and \textit{Soft State} when the flux is at a similar level to the two exotic variability classes. Considering the hinted slim disk nature in SIMS (see Section~\ref{discussion:Eddington}), it is plausible that the corona producing hard X-ray photons cannot illuminate the outer disk with a slim inner disk (scale height can approach $H/R\sim1$ near the Eddington limit) in the way. We note that studying the role of winds in both \igr\ and \grs\ might be important to understand the physical mechanism producing heartbeats (see the discussion in \citealp{2011ApJ...737...69N}).

With regard to the emission line, we note that previous phase-resolved spectroscopy analysis has detected an iron emission line in \textit{Class $\rho$} of \grs\ (the traditional heartbeat class; \citealp{2011ApJ...737...69N,2016ApJ...833..165Z}). The absence of an iron emission line from the reflection in the exotic states in this work could be an ionization effect.}

\subsubsection{Future directions}

{\color{black}The 2022 outburst of \igr\ is rich in phenomenology, and provides us a way of connecting the more typical BHXB evolution to systems that exhibit more exotic variability patterns, but there is clearly more work needed. {\color{black}Besides the spectral-timing analysis methods presented in this work, phase-resolved spectroscopy, and some novel timing methods such as recurrence analysis could shed light on the heartbeat nature. It is worth noting that a deterministic (rather than stochastic) nature of variability resulting from nonlinear dynamics was found in both \grs\ and \igr\ \citep{2016A&A...586A.143S}.}

{\color{black}Moreover, the} question remains whether \igr\ and \grs\ really are outliers, or perhaps, exotic variability is less exotic than we thought. {\color{black}For instance, we found that at the end of \textit{SIMS}, there were some flares and modulations that are signs of emerging exotic variability; and some data in \textit{Class V} could only be identified by the non-zero-centered noise component but not the lightcurves directly due to the low level of exotic variability. This dataset hints at a rethinking of the canonical state identification of BHXBs that is in general based on the disk and corona contribution to the spectrum and timing features including the total level of variability and the QPO properties; perhaps a second dimension perpendicular to the canonical thinking is the level of exotic or non-stochastic variability. It remains an open question if there is a hard boundary between `normal' and `exotic', and if so, where that boundary should be drawn.
In other BHXBs, we notice} exotic variability that is similar to \textit{Class V} was discovered in \textit{optical} in GX~339--4, an archetypical BHXB \citep{1982A&A...109L...1M}, and some `flip-flops' were observed in X-ray in its soft state \citep{1991ApJ...383..784M,2022MNRAS.513.4308L}. {\color{black}We note that `flip-flops' are usually associated with state transitions between HIMS and SIMS and corresponding QPO type transitions (Types C and B), which have also been found in XTE~J1859+226 \citep{2004A&A...426..587C} and MAXI~J1348--630 \citep{2021MNRAS.505.3823Z}.} In \citet{2011MNRAS.414.2186J}, the authors searched for heartbeat-like variability using both the lightcurves and any presence of the QPO at a frequency $<0.1$~Hz, and found several candidates showing heartbeats.
A systematic search for exotic variability in multiwavelength lightcurves of both BHXBs and neutron star X-ray binaries will help us understand if this behavior is more common than we realized. }{\color{black}The sources we should focus more on are those that reached a high Eddington ratio (XTE~J1550--564, 4U~1543--47), or have a long orbital period (XTE~J0421+560, 1E~1740.7--2942, GRS~1758--258), or have shown interesting variability in the past (GX~339--4, XTE~J1859+226, MAXI~J1348--630).}


\section{Summary} \label{summary}

We have presented the spectral-timing properties of \igr, the fainter heartbeat BHXB, in its 2022 outburst, using {\color{black}a legacy dataset of} NICER, NuSTAR, and Chandra. By aggregating results using spectral-timing tools including the lightcurves, the PSDs, and the flux-energy spectra, our major findings are as follows: 

\begin{enumerate}
\item{{\color{black}}\igr\ {\color{black}first went} through {\color{black}a normal} hard state and intermediate state. After that, instead of {\color{black}entering} a traditional soft state with a stable accretion disk, the flux decreased, and some sort of instability {\color{black}set in}, producing heartbeat-like exotic variability in the lightcurve. Then, \igr\ {\color{black}transitioned to an exotic soft state where we identify two types of heartbeat-like variabilities (\textit{Class V} and a new \textit{Class X}).} {\color{black}Eventually \igr\ went into an intermediate state approaching the initial hard state.}}
\item{{\color{black}\igr\ shares some variability classes with \grs\ that are not seen in other BHXBs, while having an outburst recurrence rate and evolution pattern in outburst similar to normal BHXBs. It can therefore be regarded as a bridge between the most peculiar BHXB \grs\ and `normal' BHXBs.}}
\item{We observe an iron emission line resulting from relativistic reflection when there is no heartbeat-like variability, and we observe absorption lines from highly ionized iron when there is heartbeat-like variability. {\color{black}This means that whether absorption lines from highly ionized iron are detected is not determined by a soft (disk-dominated) spectral state alone, but rather is determined by the presence of exotic variabilities; in a soft spectral state, absorption lines are only detected along with exotic variabilities.} The absorber {\color{black}has an outflow velocity $<0.01c$}, consistent with thermally-driven outflows originating from the outer disk.}
\item{{\color{black}We find that although a super-Eddington luminosity in \igr\ cannot be ruled out at this point, the luminosity cannot be the only driver of exotic variability because exotic variability is not always present at the same luminosity.
}}
\item{\color{black}If \igr\ reached only sub-Eddington luminosities, it remains a mystery what factor sets it apart from the other BHXBs to exhibit heartbeat-like variabilities. Some potential hypotheses involve a large disk size (although the exact mechanism is unclear and might actually relate back to a high Eddington ratio), or disk tearing. In the future, we should systematically search for exotic variability in multiwavelength lightcurves because it might be more common than we realize, and some physical mechanism might be shared.}
\end{enumerate}

\vspace{1cm}
We thank Keith Gendreau, Zaven Arzoumanian, Elizabeth Ferrara, Karl Forster, and Pat Slane for scheduling and performing the observations. 
JW acknowledges support from the NASA~FINESST Graduate Fellowship, under grant 80NSSC22K1596. JW thanks Gibwa Musoke, Megan Masterson, Mason Ng, Navin Sridhar, and Yerong Xu for useful discussions. 
JW, EK, JAG, GM, and ML acknowledge support from NASA~ADAP grant 80NSSC17K0515. AI and DA acknowledge support from the Royal Society. {\color{black}MK acknowledges support from an NWO Spinoza prize.}

\appendix

\setcounter{figure}{0}
\renewcommand{\thefigure}{A\arabic{figure}}

\setcounter{table}{0}
\renewcommand{\thetable}{A\arabic{table}}

\section{Lightcurves at the end of SIMS and Transition to Class V}

\begin{figure*}
\centering
\includegraphics[width=1.0\linewidth]{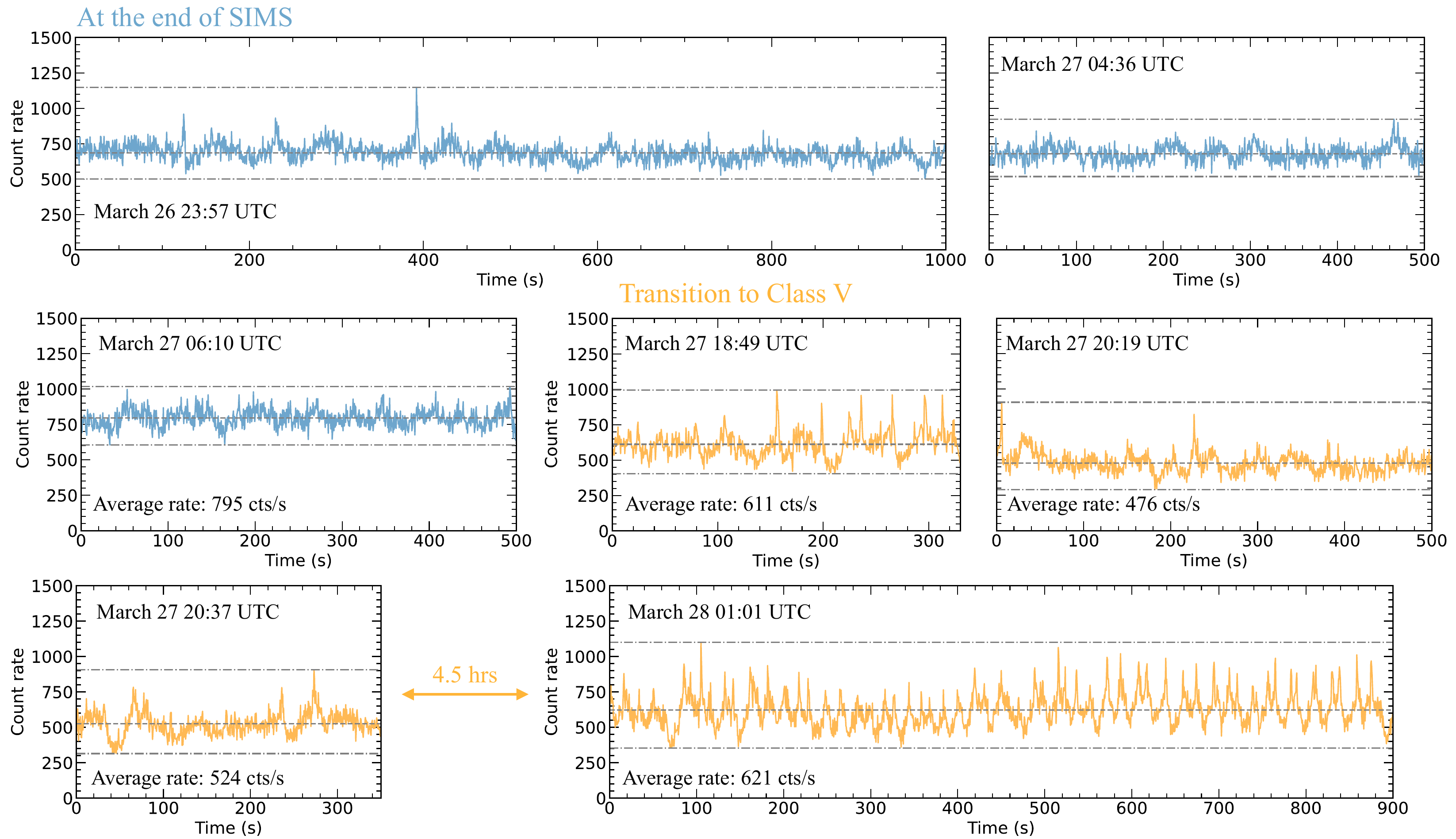}
\caption{{\color{black}The NICER lightcurves at the end of \textit{SIMS} (NICER ObsIDs 4618020402 and 5202630108) and during the \textit{Transition to Class V} (NICER ObsIDs 5202630108 and 5202630109). The dashed line represents the average count rate, and the dash-dotted lines mark the minimum and maximum count rate in the lightcurves shown. The count rate is measured in 0.3--12~keV with NICER normalized for 52 FPMs, with time bins of 0.5~s. The time in each panel marks the start time of the lightcurve. }}
\label{fig:lc_transition}
\end{figure*}

{\color{black}In Fig.~\ref{fig:lc_transition}, we show the NICER lightcurves at the end of \textit{SIMS} and during the \textit{Transition to Class V}, to show in more detail how this transitional phase proceeded. Starting from March 26, we started to observe some flares and modulations that are signs of emerging exotic variability. The more quantitative evidence for this is the peak at low frequencies $\sim0.016$~Hz in the PSDs (see Fig.~\ref{fig:psd}b). The spectral and timing features of the data remained very similar until March 27 when we compared the lightcurve at 18:49 UTC to 06:10 UTC. The average source count rate decreased from $\sim800$~counts/sec to $\sim600$~counts/sec, with stronger flares and more exotic variabilities showing up. The source count rate continued decreasing to $\sim500$~counts/sec on March 27 20:37 UTC. Then, in only 4.5~hours, the source fully transitioned into \textit{Class V} at March 28 01:01 UTC, with a slightly higher average count rate of $\sim620$~counts/sec and much stronger and coherent exotic variability. }

\section{Reflection spectroscopy parameters}\label{appendix:reflection}

As we present in Section~\ref{sect:detailed_spec}, there are relativistic reflection signatures when there is no heartbeat-like exotic variability in the lightcurves. We try to model them with a full reflection model. The parameters in the reflection model are set up as follows. We use the reflection flavor \texttt{relxilllpCp} assuming a lamppost geometry of the corona. We use a density gradient of a standard $\alpha$ disk (\texttt{iongrad\_type}$=2$), and we take into account the returning radiation (\texttt{switch\_returnrad}$=1$). The BH spin parameter $a_*$ is fixed at the maximal 0.998. The global parameters including the Galactic column density $N_{\rm H}$, the inclination angle, and the iron abundance are tied between epochs. We also add a phenomenological model \texttt{gabs} to account for residual calibration feature in the NICER spectra at the Si K$\alpha$ line energy of 1.74~keV. The line energy is fixed, while the width and depth are free to vary among epochs, allowing potential attitude dependency. However, we find the \textit{Soft State} spectrum drives the iron abundance to be very high at $\sim5$, while the other epochs showing reflection can be fitted well with a much more physical solar abundance. We thus untie the iron abundance and inclination in the \textit{Soft State} from the other states and fix the iron abundance at 1 for the other epochs. The disk electron density is difficult to constrain in the \textit{Hard State} and \textit{HIMS}, and therefore is fixed at $\log N_{\rm e}=18$. The electron temperature of the corona cannot be constrained with NICER alone in the \textit{Hard State}, \textit{HIMS}, and \textit{Transition to Class V}, so it is fixed at 100~keV in those epochs. {\color{black}For \textit{Class X} with the highest disk fraction among the 8 epochs, it is difficult to constrain the photon index when we use a NuSTAR band of 3--20~keV. We, therefore, set its lower limit at 2.3, which is the lowest measured for the other epochs besides the \textit{Hard State} and \textit{HIMS}.}

We have discussed the global parameters that are constrained with more confidence and the properties of the disk and corona from the continuum shape in Section~\ref{sect:detailed_spec}. {\color{black}We notice \igr\ is a peculiar BHXB with sometimes a very soft spectrum (dominated by the very hot disk), the assumptions in the reflection model that we use are not guaranteed to hold (e.g., the low-energy break of the Comptonization component, the assumed geometrically thin disk, and we do not account for emission from the plunging region; see e.g., \citealp{2018ApJ...855..120T,fabian2020soft}). As a consequence, the parameters constrained with reflection spectroscopy need to be taken with caveats. }This is especially true for the reflection model included for the \textit{Soft State}, where both the iron emission line and the Compton hump are weak, and a super-solar iron abundance is required. {\color{black}For reflection signatures seen in a traditional soft state, it has been found that the 
irradiation can be dominated by returning radiation from the innermost thermal disk's emission bent by strong gravity \citep{2021ApJ...909..146C}. We choose the same reflection model flavor as the other epochs for a purpose of consistency, but the most appropriate model should be carefully tested in future work.} Therefore, we counsel extra caution interpreting the parameters in the reflection model in the \textit{Soft State}, including the coronal height, inner disk radius, ionization parameter, and the disk electron density. 

The coronal height is low at $\lesssim10$~$R_{\rm g}$ in the \textit{Hard State}, \textit{HIMS}, and \textit{IMS Return}, and is {\color{black}not well constrained at} {\color{black}$100^{+210}_{-90}$}~$R_{\rm g}$ in the \textit{SIMS}. We note this {\color{black}does not formally contradict previous findings from reverberation mapping that the corona expands vertically in the intermediate state, which however} usually starts already in the HIMS \citep{wang2021disk,2021A&A...654A..14D, 2022ApJ...930...18W}. The inner edge of the disk $R_{\rm in}$ is in the range of 4--7~$R_{\rm g}$ {\color{black}($R_{\rm in}$ is in units of the ISCO radius in Table~\ref{tab:spec_pars}, and to convert to $R_{\rm g}$ units, a factor of 1.23 corresponding to the assumed maximal $a_*$ should be multiplied)}. If the disk is not truncated (i.e., $R_{\rm in}=R_{\rm ISCO}$ where $R_{\rm ISCO}$ is the innermost stable circular orbit radius that decreases monotonically with $a_*$), this corresponds to relatively low $a_*$ between -0.3 and 0.6. This is consistent with the result of $-0.13<a_*<0.27$ from reflection spectroscopy using NuSTAR data in the 2016 outburst \citep{2018MNRAS.478.4837W}.

\begin{figure}[htb!]
\centering
\includegraphics[width=1.\linewidth]{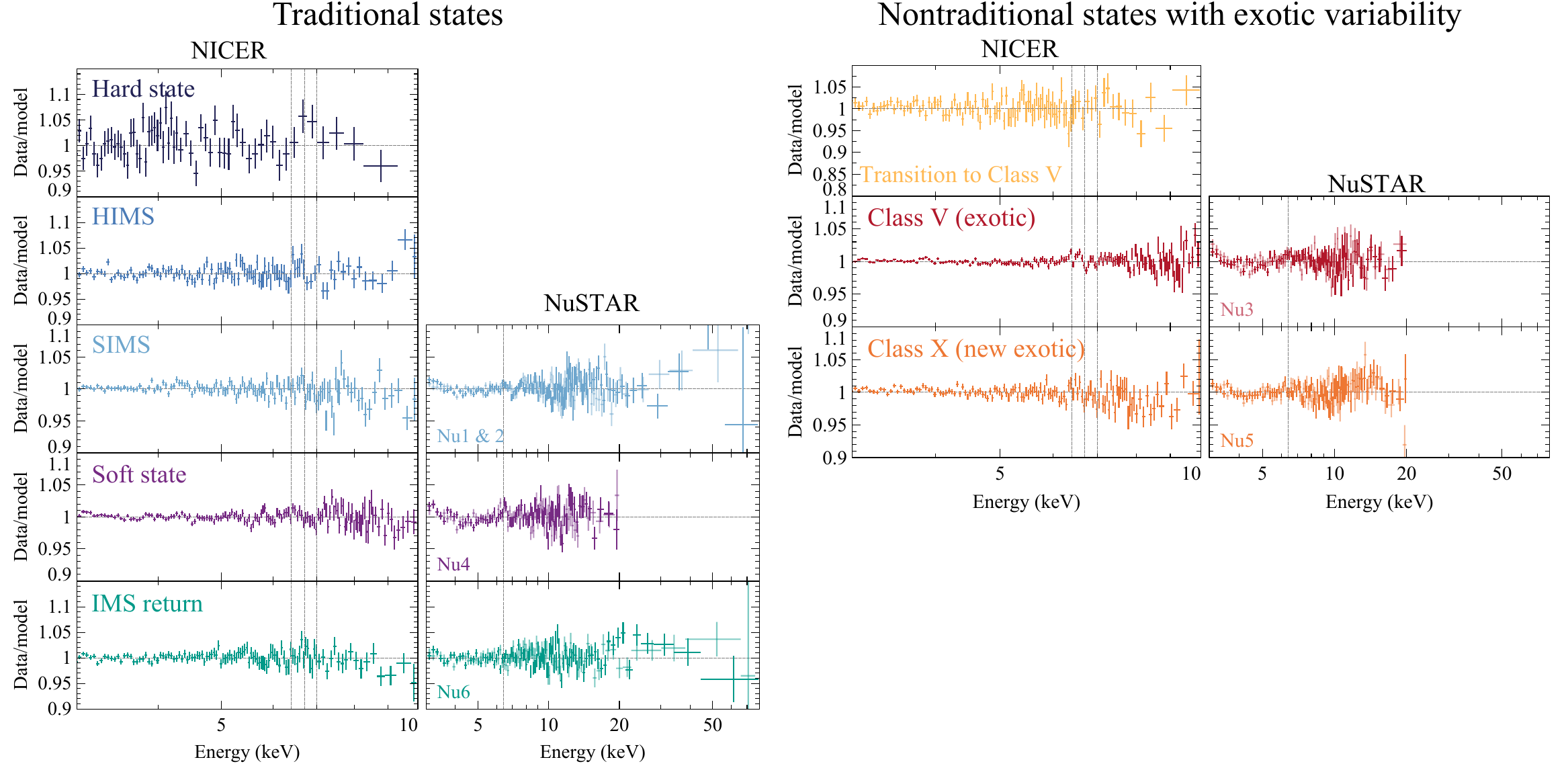}
\caption{{\color{black}The data-to-model ratio of the final model {\color{black}including the reflection and absorption lines (see Section~\ref{sect:detailed_spec} and Appendix~\ref{appendix:reflection} for more details).} (\textit{Left}) Relativistic reflection is added to the traditional states {\color{black}with stochastic variability}. (\textit{Right}) Absorption lines are added to the nontraditional states {\color{black}with exotic variability}. The dashed lines indicate Fe K$\alpha$ at 6.4~keV, {\color{black}He}-like Fe XXV at 6.7~keV, and {\color{black}H}-like Fe XXVI at 6.97~keV in the NICER spectra, and 6.4~keV in NuSTAR spectra.} {\color{black}The number after `Nu' indicates the index of the NuSTAR observation in chronological order.}}
\label{fig:ra_final}
\end{figure}

\begin{table*}[htb!]
\caption{Best fit parameters when fitting the time-averaged flux-energy spectra with the final model \texttt{crabcorr*tbabs*(diskbb+nthcomp+relxilllpCp+gauss+gauss+gauss)*gabs}.  \label{tab:spec_pars}}
\scalebox{0.8}{
\hspace{-3cm}
\begin{tabular}{cccccccccc}\hline \hline
Parameter & Epoch 1&Epoch 2 &Epoch 3&Epoch 4&Epoch 5&Epoch 6&Epoch 7&Epoch 8\\
& \textit{Hard State} & \textit{HIMS} & \textit{SIMS} & \textit{Transition to} & \textit{Class V} & \textit{Soft State} & \textit{Class X}  & \textit{IMS Return}\\
& &  &  & \textit{Class V} & (exotic) &  & (new exotic) & \\
NuSTAR Obs.&  &  & Nu1+2 &  & Nu3 & Nu4 & Nu5 & Nu6\\
\hline
$N_{\rm H}$&\multicolumn{8}{c}{$(1.537\pm0.002)\times 10^{22}$cm$^{-2}$}\\
$a_*$&\multicolumn{8}{c}{$0.998$ (f)}\\
$i$ (degrees)&\multicolumn{8}{c}{$24\pm4$ } \\
$A_{\rm Fe}$&\multicolumn{8}{c}{$1$ (f)}& \\
$T_{\rm in}$ (keV)& $0.20\pm0.02$ & $0.61^{+0.07}_{-0.02}$ & $1.463^{+0.005}_{-0.004}$ & $1.656^{+0.010}_{-0.020}$ & $1.644^{+0.005}_{-0.003}$& $1.694^{+0.013}_{-0.016}$ & $1.562^{+0.003}_{-0.004}$ & $1.028^{+0.015}_{-0.008}$\\
$N_{\rm diskbb}$ &$(1.2^{+1.2}_{-0.6})\times10^4$  & $250^{+40}_{-30}$  & $43.0\pm0.4$ & $19.4^{+2.2}_{-0.8}$  & $23.8^{+0.3}_{-1.6}$ & $27.6\pm0.2$ & $34.6\pm0.2$ & $60\pm2$ \\
$\Gamma$& $1.60\pm0.02$ & $2.033\pm0.013$ & $2.782^{+0.012}_{-0.015}$ & $3.4^{(p)}_{-0.6}$ & $2.88^{+0.33}_{-0.02}$& $3.00^{+0.08}_{-0.12}$ & $2.30^{+0.02}_{(p)}$ & $2.35^{+0.03}_{-0.04}$ \\
$kT_{\rm e}$ (keV)&$100$ (f)  & $100$ (f) & $>182$ (p) & $100$ (f)  & $5.0^{+0.3}_{(p)}$ & $5.3^{+4.8}_{-0.1}$ & $10.5^{+1.1}_{-1.3}$ & $59^{+23}_{-11}$ \\
$N_{\rm nthcomp}$ &$<0.02$  & $0.14^{+0.09}_{-0.03}$  & $0.23\pm0.01$ & $0.14^{+0.03}_{-0.04}$  & $0.250^{+0.012}_{-0.011}$ & $0.19\pm0.01$ & $0.120\pm0.003$ & $0.23\pm0.02$ \\
$h$ ($R_{\rm g}$)&  $5.0^{+1.2}_{(p)}$ & $9^{+12}_{-4 (p)}$ & $170^{+230}_{-100}$ & -& - & $40^{+10}_{-20}$ & - & $5^{+2}_{(p)}$\\
$R_{\rm in}$ ($R_{\rm ISCO}$) & $3.1^{+2.6}_{-0.9}$ & $5.6^{+1.5}_{-0.8}$ & $1.0^{+3.6}_{(p)}$ & - & - & $1.3\pm0.1$& - & $6.2\pm1.6$ \\
$\log \xi$ (erg$\cdot$cm$\cdot$s$^{-1}$) & $2.91^{+1.13}_{-0.17}$ & $3.32^{+0.21}_{-0.12}$ & $0.7^{+1.2}_{-0.2}$ & - & - & $0.84^{+0.51}_{-0.01}$ & - & $3.08^{+0.21}_{-0.19}$ \\
$\log N_{\rm e}$ (cm$^{-3}$) & $18$ (f) & $18$ (f) & $16.8^{+0.9}_{-0.3}$ & - & - & $19.8\pm0.1$ & - & $15.4^{+1.3}_{-0.4 (p)}$ \\
N$_{\rm relxill}$ ($10^{-3}$)& $15^{+3}_{-4}$ & $9\pm3$ & $12\pm2$ & - & - & $0.08\pm0.02$& - & $8.6^{+1.0}_{-1.4}$ \\
\hline
$E_{\rm gauss, 1}$ (keV)& - & - & - & $6.66^{+0.02}_{-0.03}$ & $6.95^{+0.04}_{-0.02}$ & - & $6.70\pm0.03$ & -\\
$\sigma_{\rm gauss, 1}$ (keV)& - & - & - & 0.02 (u) & 0.02 (u) & - &$0.02^{+0.03}_{(p)}$ & -\\
$N_{\rm gauss, 1}$ ($10^{-4}$)& - & - & - & $6.4^{+1.9}_{-1.6}$& $1.0\pm0.3$& - & $1.7\pm0.5$& -\\
$\rm EW_{1}$ (eV)& - & - & - & $16^{+15}_{-2}$& $2.0^{+2.4}_{-0.1}$& - & $6\pm3$& -\\
Significance level$_1$& - & - & - & $6\sigma$& $5\sigma$& - & $5\sigma$& -\\
$E_{\rm gauss, 2}$ (keV)& - & - & - & $7.01\pm0.04$ & - & - & $6.97\pm0.04$ & -\\
$\sigma_{\rm gauss, 2}$ (keV)& - & - & - & $0.02$ (u) & - & - & 0.02 (u)& -\\
$N_{\rm gauss, 2}$ ($10^{-4}$)& - & - & - & $4.0\pm1.7$ & - & - & $1.2\pm0.5$ & -\\
$\rm EW_{2}$ (eV)& - & - & - & $16^{+10}_{-9}$ & - & - & $5\pm3$ & -\\
Significance level$_2$& - & - & - & $3\sigma$ & - & - & $3\sigma$ & -\\
$E_{\rm gauss, 3}$ (keV)& - & - & - & $7.78\pm0.07$ & - & -& -& -\\
$\sigma_{\rm gauss, 3}$ (keV)& - & - & - & $0.05$ (u) & - & -& -& -\\
$N_{\rm gauss, 3}$ ($10^{-4}$)& - & - & - & $4.2^{+1.9}_{-2.0}$ & - & -& - & -\\
$\rm EW_{3}$ (eV)& - & - & - & $32^{+6}_{-23}$ & - & -& - & -\\
Significance level$_3$& - & - & - & $3\sigma$ & - & -& - & -\\
\hline
$E_{\rm gabs}$ (keV)&\multicolumn{8}{c}{$1.74$ (f)}\\ 
$\sigma$ (keV)&  $>0.03$ & $0.05^{+0.03}_{-0.02}$ & $0.05\pm0.03$ & $<0.07$ & $>0.08$& $0.04\pm0.03$ & $<0.06$& $>0.05$ \\
Strength &  $-0.007^{+0.004}_{-0.005}$ & $-0.005\pm0.002$ & $-0.005\pm0.002$ & $-0.004\pm0.002$ & $0.002^{+0.002}_{-0.004}$& $-0.005^{+0.001}_{-0.002}$ & $-0.002\pm0.001$& $-0.009\pm0.002$ \\
\hline
$\Delta\Gamma_{\rm NICER}$ & \multicolumn{8}{c}{0 (f)} \\
$N_{\rm NICER}$ & \multicolumn{8}{c}{1 (f)} \\
$\Delta\Gamma_{\rm FPMA}$& - & - & $0.098^{+0.004}_{-0.006}$ & - & $0.051\pm0.005$ & $0.129^{+0.006}_{-0.003}$ & $0.139\pm0.007$ & $0.097\pm0.004$\\
$N_{\rm FPMA}$ & - & - & $1.176^{+0.010}_{-0.004}$ & - & $1.315\pm0.012$ & $1.146^{+0.012}_{-0.006}$ & $1.314\pm0.014$ & $1.619\pm0.011$\\
$\Delta\Gamma_{\rm FPMB}$& - & - & $0.098^{+0.004}_{-0.006}$ & - & $0.048\pm0.006$ & $0.111\pm0.006$ & $0.138\pm0.007$ & $0.096\pm0.004$\\
$N_{\rm FPMB}$ & - & - & $1.160^{+0.005}_{-0.009}$ & - & $1.289\pm0.012$ & $1.100\pm0.012$ & $1.290\pm0.014$ & $1.562^{+0.011}_{-0.015}$\\
\hline
Disk fraction & $0.042^{+0.014}_{-0.022}$\% & $5.0\pm0.5$\% & $52.8^{+0.8}_{-1.4}$\% & $61^{+5}_{-22}$\% & $49.8^{+0.3}_{-0.7}$\% & $50.9^{+2.9}_{-0.4}$\% & $65.8^{+0.7}_{-0.6}$\% & $22.2^{+0.9}_{-0.6}$\%\\
Flux ($10^{-9}$ erg cm$^{-2}$ s$^{-1}$)& $1.51^{+0.05}_{-0.03}$ & $3.25\pm0.02$ & $5.044^{+0.005}_{-0.007}$ & $3.45^{+0.04}_{-0.03}$ & $4.962^{+0.003}_{-0.004}$ & $5.140^{+0.005}_{-0.006}$ & $4.401^{+0.005}_{-0.006}$ & $3.038\pm0.008$\\
\hline
$\chi^2$/d.o.f.& \multicolumn{8}{c}{$4516/3857=1.17$}\\
\hline
\end{tabular}
}
\\
\raggedright{\textbf{Notes.} \\
Errors are at 90\% confidence level and statistical only. The index $(f)$ means the parameter is fixed, $(p)$ means that the uncertainty is pegged at the bound allowed for the parameter, and $(u)$ indicates the parameter is unconstrained. The line widths of the absorption lines modeled by \texttt{gaussian} are limited in the range of 0.02 to 0.05~keV. The line width is set to have an upper limit of 0.1~keV in the phenomenological \texttt{gabs} model. 
The flux and disk fraction are both measured using \texttt{cflux} in \texttt{XSPEC} in the 2--20 keV band, and the total flux is unabsorbed flux. 
The equivalent width is calculated using the \texttt{XSPEC} command \texttt{eqwidth}. {\color{black}Even though the model provides a good fit for the spectra, the exact values of parameters constrained from reflection models that are constrained mostly with the details in the iron line need to be taken with caveats (e.g., the coronal height, black hole spin, inclination, etc.), because the assumptions in the reflection model used are not guaranteed to hold.}}
\end{table*}

\bibliographystyle{apj}
\bibliography{draftv5_arXiv}

\begin{thebibliography}{}
\expandafter\ifx\csname natexlab\endcsname\relax\def\natexlab#1{#1}\fi

\bibitem[{{Abramowicz} {et~al.}(1988){Abramowicz}, {Czerny}, {Lasota}, \& {Szuszkiewicz}}]{1988ApJ...332..646A}
{Abramowicz}, M.~A., {Czerny}, B., {Lasota}, J.~P., \& {Szuszkiewicz}, E. 1988, \apj, 332, 646

\bibitem[{{Abramowicz} {et~al.}(2010){Abramowicz}, {Jaroszy{\'n}ski}, {Kato}, {Lasota}, {R{\'o}{\.z}a{\'n}ska}, \& {S{\k{a}}dowski}}]{2010A&A...521A..15A}
{Abramowicz}, M.~A., {Jaroszy{\'n}ski}, M., {Kato}, S., {et~al.} 2010, \aap, 521, A15

\bibitem[{{Altamirano} \& {Belloni}(2012)}]{2012ApJ...747L...4A}
{Altamirano}, D., \& {Belloni}, T. 2012, \apjl, 747, L4

\bibitem[{{Altamirano} {et~al.}(2011){Altamirano}, {Belloni}, {Linares}, {van der Klis}, {Wijnands}, {Curran}, {Kalamkar}, {Stiele}, {Motta}, {Mu{\~n}oz-Darias}, {Casella}, \& {Krimm}}]{altamirano2011faint}
{Altamirano}, D., {Belloni}, T., {Linares}, M., {et~al.} 2011, \apjl, 742, L17

\bibitem[{{Arnaud}(1996)}]{arn96}
{Arnaud}, K.~A. 1996, in Astronomical Society of the Pacific Conference Series, Vol. 101, Astronomical Data Analysis Software and Systems V, ed. G.~H. {Jacoby} \& J.~{Barnes}, 17

\bibitem[{{Bagnoli} \& {in't Zand}(2015)}]{2015MNRAS.450L..52B}
{Bagnoli}, T., \& {in't Zand}, J.~J.~M. 2015, \mnras, 450, L52

\bibitem[{{Belloni} \& {Hasinger}(1990)}]{1990A&A...227L..33B}
{Belloni}, T., \& {Hasinger}, G. 1990, \aap, 227, L33

\bibitem[{{Belloni} {et~al.}(2000){Belloni}, {Klein-Wolt}, {M{\'e}ndez}, {van der Klis}, \& {van Paradijs}}]{2000A&A...355..271B}
{Belloni}, T., {Klein-Wolt}, M., {M{\'e}ndez}, M., {van der Klis}, M., \& {van Paradijs}, J. 2000, \aap, 355, 271

\bibitem[{{Belloni} {et~al.}(1997){Belloni}, {M{\'e}ndez}, {King}, {van der Klis}, \& {van Paradijs}}]{1997ApJ...488L.109B}
{Belloni}, T., {M{\'e}ndez}, M., {King}, A.~R., {van der Klis}, M., \& {van Paradijs}, J. 1997, \apjl, 488, L109

\bibitem[{{Belloni} {et~al.}(2011){Belloni}, {Motta}, \& {Mu{\~n}oz-Darias}}]{2011BASI...39..409B}
{Belloni}, T.~M., {Motta}, S.~E., \& {Mu{\~n}oz-Darias}, T. 2011, Bulletin of the Astronomical Society of India, 39, 409

\bibitem[{{Bland-Hawthorn} \& {Gerhard}(2016)}]{2016ARA&A..54..529B}
{Bland-Hawthorn}, J., \& {Gerhard}, O. 2016, \araa, 54, 529

\bibitem[{Bogdanov(2019)}]{bogdanov2019constraining}
Bogdanov, S. 2019, in American Astronomical Society Meeting Abstracts\# 233, Vol. 233

\bibitem[{{Canizares} {et~al.}(2005){Canizares}, {Davis}, {Dewey}, {Flanagan}, {Galton}, {Huenemoerder}, {Ishibashi}, {Markert}, {Marshall}, {McGuirk}, {Schattenburg}, {Schulz}, {Smith}, \& {Wise}}]{2005PASP..117.1144C}
{Canizares}, C.~R., {Davis}, J.~E., {Dewey}, D., {et~al.} 2005, \pasp, 117, 1144

\bibitem[{{Capitanio} {et~al.}(2006){Capitanio}, {Bazzano}, {Ubertini}, {Zdziarski}, {Bird}, {De Cesare}, {Dean}, {Stephen}, \& {Tarana}}]{2006ApJ...643..376C}
{Capitanio}, F., {Bazzano}, A., {Ubertini}, P., {et~al.} 2006, \apj, 643, 376

\bibitem[{{Capitanio} {et~al.}(2009){Capitanio}, {Giroletti}, {Molina}, {Bazzano}, {Tarana}, {Kennea}, {Dean}, {Hill}, {Tavani}, \& {Ubertini}}]{2009ApJ...690.1621C}
{Capitanio}, F., {Giroletti}, M., {Molina}, M., {et~al.} 2009, \apj, 690, 1621

\bibitem[{{Casella} {et~al.}(2004){Casella}, {Belloni}, {Homan}, \& {Stella}}]{2004A&A...426..587C}
{Casella}, P., {Belloni}, T., {Homan}, J., \& {Stella}, L. 2004, \aap, 426, 587

\bibitem[{{Castro-Tirado} {et~al.}(1992){Castro-Tirado}, {Brandt}, \& {Lund}}]{1992IAUC.5590....2C}
{Castro-Tirado}, A.~J., {Brandt}, S., \& {Lund}, N. 1992, \iaucirc, 5590, 2

\bibitem[{{Connors} {et~al.}(2021){Connors}, {Garc{\'\i}a}, {Tomsick}, {Hare}, {Dauser}, {Grinberg}, {Steiner}, {Mastroserio}, {Sridhar}, {Fabian}, {Jiang}, {Parker}, {Harrison}, \& {Kallman}}]{2021ApJ...909..146C}
{Connors}, R. M.~T., {Garc{\'\i}a}, J.~A., {Tomsick}, J., {et~al.} 2021, \apj, 909, 146

\bibitem[{{Court} {et~al.}(2017){Court}, {Altamirano}, {Pereyra}, {Boon}, {Yamaoka}, {Belloni}, {Wijnands}, \& {Pahari}}]{2017MNRAS.468.4748C}
{Court}, J.~M.~C., {Altamirano}, D., {Pereyra}, M., {et~al.} 2017, \mnras, 468, 4748

\bibitem[{{Dauser} {et~al.}(2022){Dauser}, {Garc{\'\i}a}, {Joyce}, {Licklederer}, {Connors}, {Ingram}, {Reynolds}, \& {Wilms}}]{2022MNRAS.514.3965D}
{Dauser}, T., {Garc{\'\i}a}, J.~A., {Joyce}, A., {et~al.} 2022, \mnras, 514, 3965

\bibitem[{{De Marco} {et~al.}(2021){De Marco}, {Zdziarski}, {Ponti}, {Migliori}, {Belloni}, {Segovia Otero}, {Dzie{\l}ak}, \& {Lai}}]{2021A&A...654A..14D}
{De Marco}, B., {Zdziarski}, A.~A., {Ponti}, G., {et~al.} 2021, \aap, 654, A14

\bibitem[{{Degenaar} {et~al.}(2014){Degenaar}, {Miller}, {Harrison}, {Kennea}, {Kouveliotou}, \& {Younes}}]{2014ApJ...796L...9D}
{Degenaar}, N., {Miller}, J.~M., {Harrison}, F.~A., {et~al.} 2014, \apjl, 796, L9

\bibitem[{Done {et~al.}(2007)Done, Gierli{\'n}ski, \& Kubota}]{done2007modelling}
Done, C., Gierli{\'n}ski, M., \& Kubota, A. 2007, The Astronomy and Astrophysics Review, 15, 1

\bibitem[{{Done} {et~al.}(2004){Done}, {Wardzi{\'n}ski}, \& {Gierli{\'n}ski}}]{done2004grs}
{Done}, C., {Wardzi{\'n}ski}, G., \& {Gierli{\'n}ski}, M. 2004, \mnras, 349, 393

\bibitem[{{Dunn} {et~al.}(2010){Dunn}, {Fender}, {K{\"o}rding}, {Belloni}, \& {Cabanac}}]{2010MNRAS.403...61D}
{Dunn}, R.~J.~H., {Fender}, R.~P., {K{\"o}rding}, E.~G., {Belloni}, T., \& {Cabanac}, C. 2010, \mnras, 403, 61

\bibitem[{{Fabian} {et~al.}(2020){Fabian}, {Buisson}, {Kosec}, {Reynolds}, {Wilkins}, {Tomsick}, {Walton}, {Gandhi}, {Altamirano}, {Arzoumanian}, {Cackett}, {Dyda}, {Garcia}, {Gendreau}, {Grefenstette}, {Homan}, {Kara}, {Ludlam}, {Miller}, \& {Steiner}}]{fabian2020soft}
{Fabian}, A.~C., {Buisson}, D.~J., {Kosec}, P., {et~al.} 2020, \mnras, 493, 5389

\bibitem[{{Fender} \& {Belloni}(2004)}]{2004ARA&A..42..317F}
{Fender}, R., \& {Belloni}, T. 2004, \araa, 42, 317

\bibitem[{{Fender} {et~al.}(1997){Fender}, {Pooley}, {Brocksopp}, \& {Newell}}]{1997MNRAS.290L..65F}
{Fender}, R.~P., {Pooley}, G.~G., {Brocksopp}, C., \& {Newell}, S.~J. 1997, \mnras, 290, L65

\bibitem[{{Finger} {et~al.}(1996){Finger}, {Koh}, {Nelson}, {Prince}, {Vaughan}, \& {Wilson}}]{1996Natur.381..291F}
{Finger}, M.~H., {Koh}, D.~T., {Nelson}, R.~W., {et~al.} 1996, \nat, 381, 291

\bibitem[{{Frank} {et~al.}(2002){Frank}, {King}, \& {Raine}}]{2002apa..book.....F}
{Frank}, J., {King}, A., \& {Raine}, D.~J. 2002, {Accretion Power in Astrophysics: Third Edition}

\bibitem[{Garc{\'\i}a {et~al.}(2014)Garc{\'\i}a, Dauser, Lohfink, Kallman, Steiner, McClintock, Brenneman, Wilms, Eikmann, Reynolds, {et~al.}}]{garcia2014_relxill}
Garc{\'\i}a, J., Dauser, T., Lohfink, A., {et~al.} 2014, The Astrophysical Journal, 782, 76

\bibitem[{Gendreau {et~al.}(2016)Gendreau, Arzoumanian, Adkins, Albert, Anders, Aylward, Baker, Balsamo, Bamford, Benegalrao, {et~al.}}]{gendreau2016neutron}
Gendreau, K.~C., Arzoumanian, Z., Adkins, P.~W., {et~al.} 2016, in Space Telescopes and Instrumentation 2016: Ultraviolet to Gamma Ray, Vol. 9905, International Society for Optics and Photonics, 99051H

\bibitem[{{Greiner} {et~al.}(2001){Greiner}, {Cuby}, \& {McCaughrean}}]{2001Natur.414..522G}
{Greiner}, J., {Cuby}, J.~G., \& {McCaughrean}, M.~J. 2001, \nat, 414, 522

\bibitem[{{Hannikainen} {et~al.}(2005){Hannikainen}, {Rodriguez}, {Vilhu}, {Hjalmarsdotter}, {Zdziarski}, {Belloni}, {Poutanen}, {Wu}, {Shaw}, {Beckmann}, {Hunstead}, {Pooley}, {Westergaard}, {Mirabel}, {Hakala}, {Castro-Tirado}, \& {Durouchoux}}]{2005A&A...435..995H}
{Hannikainen}, D.~C., {Rodriguez}, J., {Vilhu}, O., {et~al.} 2005, \aap, 435, 995

\bibitem[{Harrison {et~al.}(2013)Harrison, Craig, Christensen, Hailey, Zhang, Boggs, Stern, Cook, Forster, Giommi, {et~al.}}]{harrison2013nuclear}
Harrison, F.~A., Craig, W.~W., Christensen, F.~E., {et~al.} 2013, The Astrophysical Journal, 770, 103

\bibitem[{{Homan} {et~al.}(2019){Homan}, {Neilsen}, {Steiner}, {Remillard}, {Altamirano}, {Gendreau}, \& {Arzoumanian}}]{2019ATel12742....1H}
{Homan}, J., {Neilsen}, J., {Steiner}, J., {et~al.} 2019, The Astronomer's Telegram, 12742, 1

\bibitem[{{Honma} {et~al.}(1991){Honma}, {Matsumoto}, \& {Kato}}]{1991PASJ...43..147H}
{Honma}, F., {Matsumoto}, R., \& {Kato}, S. 1991, \pasj, 43, 147

\bibitem[{Ingram \& Motta(2019)}]{ingram2019review}
Ingram, A.~R., \& Motta, S.~E. 2019, New Astronomy Reviews, 85, 101524

\bibitem[{{Iyer} {et~al.}(2015){Iyer}, {Nandi}, \& {Mandal}}]{2015ApJ...807..108I}
{Iyer}, N., {Nandi}, A., \& {Mandal}, S. 2015, \apj, 807, 108

\bibitem[{{Janiuk} \& {Czerny}(2011)}]{2011MNRAS.414.2186J}
{Janiuk}, A., \& {Czerny}, B. 2011, \mnras, 414, 2186

\bibitem[{{Janiuk} {et~al.}(2000){Janiuk}, {Czerny}, \& {Siemiginowska}}]{2000ApJ...542L..33J}
{Janiuk}, A., {Czerny}, B., \& {Siemiginowska}, A. 2000, \apjl, 542, L33

\bibitem[{{Janiuk} {et~al.}(2002){Janiuk}, {Czerny}, \& {Siemiginowska}}]{2002ApJ...576..908J}
---. 2002, \apj, 576, 908

\bibitem[{{Janiuk} {et~al.}(2015){Janiuk}, {Grzedzielski}, {Capitanio}, \& {Bianchi}}]{janiuk2015interplay}
{Janiuk}, A., {Grzedzielski}, M., {Capitanio}, F., \& {Bianchi}, S. 2015, \aap, 574, A92

\bibitem[{{Kalemci} {et~al.}(2022){Kalemci}, {Kara}, \& {Tomsick}}]{2022arXiv220614410K}
{Kalemci}, E., {Kara}, E., \& {Tomsick}, J.~A. 2022, arXiv e-prints, arXiv:2206.14410

\bibitem[{{Kimura} {et~al.}(2016){Kimura}, {Isogai}, {Kato}, {Ueda}, {Nakahira}, {Shidatsu}, {Enoto}, {Hori}, {Nogami}, {Littlefield}, {Ishioka}, {Chen}, {King}, {Wen}, {Wang}, {Lehner}, {Schwamb}, {Wang}, {Zhang}, {Alcock}, {Axelrod}, {Bianco}, {Byun}, {Chen}, {Cook}, {Kim}, {Lee}, {Marshall}, {Pavlenko}, {Antonyuk}, {Antonyuk}, {Pit}, {Sosnovskij}, {Babina}, {Baklanov}, {Pozanenko}, {Mazaeva}, {Schmalz}, {Reva}, {Belan}, {Inasaridze}, {Tungalag}, {Volnova}, {Molotov}, {de Miguel}, {Kasai}, {Stein}, {Dubovsky}, {Kiyota}, {Miller}, {Richmond}, {Goff}, {Andreev}, {Takahashi}, {Kojiguchi}, {Sugiura}, {Takeda}, {Yamada}, {Matsumoto}, {James}, {Pickard}, {Tordai}, {Maeda}, {Ruiz}, {Miyashita}, {Cook}, {Imada}, \& {Uemura}}]{2016Natur.529...54K}
{Kimura}, M., {Isogai}, K., {Kato}, T., {et~al.} 2016, \nat, 529, 54

\bibitem[{{King} {et~al.}(2012){King}, {Miller}, {Raymond}, {Fabian}, {Reynolds}, {Kallman}, {Maitra}, {Cackett}, \& {Rupen}}]{king2012extreme}
{King}, A.~L., {Miller}, J.~M., {Raymond}, J., {et~al.} 2012, \apjl, 746, L20

\bibitem[{{Klein-Wolt} {et~al.}(2002){Klein-Wolt}, {Fender}, {Pooley}, {Belloni}, {Migliari}, {Morgan}, \& {van der Klis}}]{2002MNRAS.331..745K}
{Klein-Wolt}, M., {Fender}, R.~P., {Pooley}, G.~G., {et~al.} 2002, \mnras, 331, 745

\bibitem[{{Kouveliotou} {et~al.}(1996){Kouveliotou}, {van Paradijs}, {Fishman}, {Briggs}, {Kommers}, {Harmon}, {Meegan}, \& {Lewin}}]{1996Natur.379..799K}
{Kouveliotou}, C., {van Paradijs}, J., {Fishman}, G.~J., {et~al.} 1996, \nat, 379, 799

\bibitem[{{Kuulkers} {et~al.}(2003){Kuulkers}, {Lutovinov}, {Parmar}, {Capitanio}, {Mowlavi}, \& {Hermsen}}]{2003ATel..149....1K}
{Kuulkers}, E., {Lutovinov}, A., {Parmar}, A., {et~al.} 2003, The Astronomer's Telegram, 149, 1

\bibitem[{{Lightman} \& {Eardley}(1974)}]{1974ApJ...187L...1L}
{Lightman}, A.~P., \& {Eardley}, D.~M. 1974, \apjl, 187, L1

\bibitem[{{Lin} {et~al.}(2009){Lin}, {Remillard}, \& {Homan}}]{2009ApJ...696.1257L}
{Lin}, D., {Remillard}, R.~A., \& {Homan}, J. 2009, \apj, 696, 1257

\bibitem[{{Liska} {et~al.}(2019){Liska}, {Tchekhovskoy}, {Ingram}, \& {van der Klis}}]{liska2019bardeen}
{Liska}, M., {Tchekhovskoy}, A., {Ingram}, A., \& {van der Klis}, M. 2019, \mnras, 487, 550

\bibitem[{{Liu} {et~al.}(2022){Liu}, {Jiang}, {Zhang}, {Bambi}, {Ji}, {Kong}, \& {Zhang}}]{2022MNRAS.513.4308L}
{Liu}, H., {Jiang}, J., {Zhang}, Z., {et~al.} 2022, \mnras, 513, 4308

\bibitem[{{Maccarone}(2003)}]{2003A&A...409..697M}
{Maccarone}, T.~J. 2003, \aap, 409, 697

\bibitem[{{Marcel} {et~al.}(2022){Marcel}, {Ferreira}, {Petrucci}, {Barnier}, {Malzac}, {Marino}, {Coriat}, {Clavel}, {Reynolds}, {Neilsen}, {Belmont}, \& {Corbel}}]{2022A&A...659A.194M}
{Marcel}, G., {Ferreira}, J., {Petrucci}, P.~O., {et~al.} 2022, \aap, 659, A194

\bibitem[{{Miller} {et~al.}(2022){Miller}, {Draghis}, {Gendreau}, \& {Arzoumanian}}]{miller2022nicer}
{Miller}, J.~M., {Draghis}, P., {Gendreau}, K., \& {Arzoumanian}, Z. 2022, The Astronomer's Telegram, 15282, 1

\bibitem[{{Miyamoto} {et~al.}(1991){Miyamoto}, {Kimura}, {Kitamoto}, {Dotani}, \& {Ebisawa}}]{1991ApJ...383..784M}
{Miyamoto}, S., {Kimura}, K., {Kitamoto}, S., {Dotani}, T., \& {Ebisawa}, K. 1991, \apj, 383, 784

\bibitem[{{M{\"o}nkk{\"o}nen} {et~al.}(2019){M{\"o}nkk{\"o}nen}, {Tsygankov}, {Mushtukov}, {Doroshenko}, {Suleimanov}, \& {Poutanen}}]{2019A&A...626A.106M}
{M{\"o}nkk{\"o}nen}, J., {Tsygankov}, S.~S., {Mushtukov}, A.~A., {et~al.} 2019, \aap, 626, A106

\bibitem[{{Morgan} {et~al.}(1997){Morgan}, {Remillard}, \& {Greiner}}]{1997ApJ...482..993M}
{Morgan}, E.~H., {Remillard}, R.~A., \& {Greiner}, J. 1997, \apj, 482, 993

\bibitem[{{Motch} {et~al.}(1982){Motch}, {Ilovaisky}, \& {Chevalier}}]{1982A&A...109L...1M}
{Motch}, C., {Ilovaisky}, S.~A., \& {Chevalier}, C. 1982, \aap, 109, L1

\bibitem[{{Motta} {et~al.}(2011){Motta}, {Mu{\~n}oz-Darias}, {Casella}, {Belloni}, \& {Homan}}]{2011MNRAS.418.2292M}
{Motta}, S., {Mu{\~n}oz-Darias}, T., {Casella}, P., {Belloni}, T., \& {Homan}, J. 2011, \mnras, 418, 2292

\bibitem[{{Motta} {et~al.}(2020){Motta}, {Marelli}, {Pintore}, {Esposito}, {Salvaterra}, {De Luca}, {Israel}, {Tiengo}, \& {Castillo}}]{2020ApJ...898..174M}
{Motta}, S.~E., {Marelli}, M., {Pintore}, F., {et~al.} 2020, \apj, 898, 174

\bibitem[{{Nayakshin} {et~al.}(2000){Nayakshin}, {Rappaport}, \& {Melia}}]{2000ApJ...535..798N}
{Nayakshin}, S., {Rappaport}, S., \& {Melia}, F. 2000, \apj, 535, 798

\bibitem[{{Negoro} {et~al.}(2018){Negoro}, {Tachibana}, {Kawai}, {Yamaoka}, {Ueda}, {Nakajima}, {Sakamaki}, {Maruyama}, {Mihara}, {Nakahira}, {Yatabe}, {Takao}, {Matsuoka}, {Sakamoto}, {Serino}, {Sugita}, {Kawakubo}, {Hashimoto}, {Yoshida}, {Sugizaki}, {Morita}, {Ueno}, {Tomida}, {Ishikawa}, {Sugawara}, {Isobe}, {Shimomukai}, {Tanimoto}, {Morita}, {Yamada}, {Tsuboi}, {Iwakiri}, {Sasaki}, {Kawai}, {Sato}, {Tsunemi}, {Yoneyama}, {Yamauchi}, {Hidaka}, {Iwahori}, {Kawamuro}, \& {Shidatsu}}]{2018ATel11828....1N}
{Negoro}, H., {Tachibana}, Y., {Kawai}, N., {et~al.} 2018, The Astronomer's Telegram, 11828, 1

\bibitem[{{Negoro} {et~al.}(2021){Negoro}, {Nakajima}, {Kobayashi}, {Mihara}, {Kawai}, {Asakura}, {Seino}, {Tamagawa}, {Matsuoka}, {Sakamoto}, {Serino}, {Sugita}, {Komachi}, {Yoshida}, {Tsuboi}, {Iwakiri}, {Kawai}, {Okamoto}, {Kitakoga}, {Shidatsu}, {Niwano}, {Hosokawa}, {Nakahira}, {Sugawara}, {Ueno}, {Tomida}, {Ishikawa}, {Tominaga}, {Nagatsuka}, {Ueda}, {Yamada}, {Ogawa}, {Setoguchi}, {Yoshitake}, {Goto}, {Uematsu}, {Tsunemi}, {Yamauchi}, {Nonaka}, {Sato}, {Hatsuda}, {Fukuoka}, {Kawamuro}, {Yamaoka}, {Kawakubo}, {Sugizaki}, \& {MAXI Team}}]{2021ATel14708....1N}
{Negoro}, H., {Nakajima}, M., {Kobayashi}, K., {et~al.} 2021, The Astronomer's Telegram, 14708, 1

\bibitem[{{Neilsen} {et~al.}(2011){Neilsen}, {Remillard}, \& {Lee}}]{2011ApJ...737...69N}
{Neilsen}, J., {Remillard}, R.~A., \& {Lee}, J.~C. 2011, \apj, 737, 69

\bibitem[{{Pahari} {et~al.}(2014){Pahari}, {Yadav}, \& {Bhattacharyya}}]{2014ApJ...783..141P}
{Pahari}, M., {Yadav}, J.~S., \& {Bhattacharyya}, S. 2014, \apj, 783, 141

\bibitem[{{Podsiadlowski} {et~al.}(2002){Podsiadlowski}, {Rappaport}, \& {Pfahl}}]{2002ApJ...565.1107P}
{Podsiadlowski}, P., {Rappaport}, S., \& {Pfahl}, E.~D. 2002, \apj, 565, 1107

\bibitem[{{Ponti} {et~al.}(2016){Ponti}, {Bianchi}, {Mu{\~n}oz-Darias}, {De}, {Fender}, \& {Merloni}}]{2016AN....337..512P}
{Ponti}, G., {Bianchi}, S., {Mu{\~n}oz-Darias}, T., {et~al.} 2016, Astronomische Nachrichten, 337, 512

\bibitem[{{Ponti} {et~al.}(2012){Ponti}, {Fender}, {Begelman}, {Dunn}, {Neilsen}, \& {Coriat}}]{2012MNRAS.422L..11P}
{Ponti}, G., {Fender}, R.~P., {Begelman}, M.~C., {et~al.} 2012, \mnras, 422, L11

\bibitem[{{Pooley} \& {Fender}(1997)}]{1997MNRAS.292..925P}
{Pooley}, G.~G., \& {Fender}, R.~P. 1997, \mnras, 292, 925

\bibitem[{{Rahoui} {et~al.}(2010){Rahoui}, {Chaty}, {Rodriguez}, {Fuchs}, {Mirabel}, \& {Pooley}}]{2010ApJ...715.1191R}
{Rahoui}, F., {Chaty}, S., {Rodriguez}, J., {et~al.} 2010, \apj, 715, 1191

\bibitem[{{Raj} \& {Nixon}(2021)}]{2021ApJ...909...82R}
{Raj}, A., \& {Nixon}, C.~J. 2021, \apj, 909, 82

\bibitem[{{Raj} {et~al.}(2021){Raj}, {Nixon}, \& {Do{\u{g}}an}}]{2021ApJ...909...81R}
{Raj}, A., {Nixon}, C.~J., \& {Do{\u{g}}an}, S. 2021, \apj, 909, 81

\bibitem[{{Reid} {et~al.}(2014{\natexlab{a}}){Reid}, {McClintock}, {Steiner}, {Steeghs}, {Remillard}, {Dhawan}, \& {Narayan}}]{2014ApJ...796....2R}
{Reid}, M.~J., {McClintock}, J.~E., {Steiner}, J.~F., {et~al.} 2014{\natexlab{a}}, \apj, 796, 2

\bibitem[{{Reid} {et~al.}(2014{\natexlab{b}}){Reid}, {McClintock}, {Steiner}, {Steeghs}, {Remillard}, {Dhawan}, \& {Narayan}}]{reid2014grs1915}
---. 2014{\natexlab{b}}, \apj, 796, 2

\bibitem[{{Reis} {et~al.}(2012){Reis}, {Miller}, {King}, \& {Reynolds}}]{reis2012igr}
{Reis}, R.~C., {Miller}, J.~M., {King}, A.~L., \& {Reynolds}, M.~T. 2012, The Astronomer's Telegram, 4382, 1

\bibitem[{{Remillard} {et~al.}(2022){Remillard}, {Loewenstein}, {Steiner}, {Prigozhin}, {LaMarr}, {Enoto}, {Gendreau}, {Arzoumanian}, {Markwardt}, {Basak}, {Stevens}, {Ray}, {Altamirano}, \& {Buisson}}]{2022AJ....163..130R}
{Remillard}, R.~A., {Loewenstein}, M., {Steiner}, J.~F., {et~al.} 2022, \aj, 163, 130

\bibitem[{{Reynolds} \& {Miller}(2011)}]{2011ApJ...734L..17R}
{Reynolds}, M.~T., \& {Miller}, J.~M. 2011, \apjl, 734, L17

\bibitem[{{Rodriguez} {et~al.}(2011){Rodriguez}, {Corbel}, {Caballero}, {Tomsick}, {Tzioumis}, {Paizis}, {Cadolle Bel}, \& {Kuulkers}}]{2011A&A...533L...4R}
{Rodriguez}, J., {Corbel}, S., {Caballero}, I., {et~al.} 2011, \aap, 533, L4

\bibitem[{{Rodriguez} {et~al.}(2003){Rodriguez}, {Corbel}, \& {Tomsick}}]{2003ApJ...595.1032R}
{Rodriguez}, J., {Corbel}, S., \& {Tomsick}, J.~A. 2003, \apj, 595, 1032

\bibitem[{Steiner {et~al.}(2010)Steiner, McClintock, Remillard, Gou, Yamada, \& Narayan}]{steiner2010constant}
Steiner, J.~F., McClintock, J.~E., Remillard, R.~A., {et~al.} 2010, The Astrophysical Journal Letters, 718, L117

\bibitem[{{Sukov{\'a}} {et~al.}(2016){Sukov{\'a}}, {Grzedzielski}, \& {Janiuk}}]{2016A&A...586A.143S}
{Sukov{\'a}}, P., {Grzedzielski}, M., \& {Janiuk}, A. 2016, \aap, 586, A143

\bibitem[{{Taylor} \& {Reynolds}(2018)}]{2018ApJ...855..120T}
{Taylor}, C., \& {Reynolds}, C.~S. 2018, \apj, 855, 120

\bibitem[{Tetarenko {et~al.}(2016)Tetarenko, Sivakoff, Heinke, \& Gladstone}]{tetarenko2016watchdog}
Tetarenko, B., Sivakoff, G., Heinke, C., \& Gladstone, J. 2016, The Astrophysical Journal Supplement Series, 222, 15

\bibitem[{Uttley {et~al.}(2014)Uttley, Cackett, Fabian, Kara, \& Wilkins}]{uttley2014x}
Uttley, P., Cackett, E., Fabian, A., Kara, E., \& Wilkins, D. 2014, The Astronomy and Astrophysics Review, 22, 72

\bibitem[{Verner {et~al.}(1996)Verner, Ferland, Korista, \& Yakovlev}]{verner1996atomic}
Verner, D., Ferland, G., Korista, K., \& Yakovlev, D. 1996, arXiv preprint astro-ph/9601009

\bibitem[{{Vincentelli} {et~al.}(2023){Vincentelli}, {Neilsen}, {Tetarenko}, {Cavecchi}, {Castro Segura}, {del Palacio}, {van den Eijnden}, {Vasilopoulos}, {Altamirano}, {Armas Padilla}, {Bailyn}, {Belloni}, {Buisson}, {Cuneo}, {Degenaar}, {Knigge}, {Long}, {Jimenez-Ibarra}, {Milburn}, {Mu{\~n}oz Darias}, {Ozbey Arabaci}, {Remillard}, \& {Russell}}]{2023arXiv230300020V}
{Vincentelli}, F.~M., {Neilsen}, J., {Tetarenko}, A.~J., {et~al.} 2023, arXiv e-prints, arXiv:2303.00020

\bibitem[{{Wang} {et~al.}(2021){Wang}, {Mastroserio}, {Kara}, {Garc{\'\i}a}, {Ingram}, {Connors}, {van der Klis}, {Dauser}, {Steiner}, {Buisson}, {Homan}, {Lucchini}, {Fabian}, {Bright}, {Fender}, {Cackett}, \& {Remillard}}]{wang2021disk}
{Wang}, J., {Mastroserio}, G., {Kara}, E., {et~al.} 2021, \apjl, 910, L3

\bibitem[{{Wang} {et~al.}(2022{\natexlab{a}}){Wang}, {Kara}, {Altamirano}, {Miller}, {Wang}, {Steiner}, {Draghis}, {Gendreau}, {Arzoumanian}, \& {Nicer Team}}]{wang2022nicer_b}
{Wang}, J., {Kara}, E., {Altamirano}, D., {et~al.} 2022{\natexlab{a}}, The Astronomer's Telegram, 15304, 1

\bibitem[{{Wang} {et~al.}(2022{\natexlab{b}}){Wang}, {Kara}, {Miller}, {Draghis}, {Gendreau}, {Arzoumanian}, {Wang}, {Altamirano}, \& {Nicer Team}}]{wang2022nicer_a}
{Wang}, J., {Kara}, E., {Miller}, J., {et~al.} 2022{\natexlab{b}}, The Astronomer's Telegram, 15287, 1

\bibitem[{{Wang} {et~al.}(2022{\natexlab{c}}){Wang}, {Kara}, {Lucchini}, {Ingram}, {van der Klis}, {Mastroserio}, {Garc{\'\i}a}, {Dauser}, {Connors}, {Fabian}, {Steiner}, {Remillard}, {Cackett}, {Uttley}, \& {Altamirano}}]{2022ApJ...930...18W}
{Wang}, J., {Kara}, E., {Lucchini}, M., {et~al.} 2022{\natexlab{c}}, \apj, 930, 18

\bibitem[{{Wang} {et~al.}(2018){Wang}, {M{\'e}ndez}, {Altamirano}, {Court}, {Beri}, \& {Cheng}}]{2018MNRAS.478.4837W}
{Wang}, Y., {M{\'e}ndez}, M., {Altamirano}, D., {et~al.} 2018, \mnras, 478, 4837

\bibitem[{{Watarai} {et~al.}(2000){Watarai}, {Fukue}, {Takeuchi}, \& {Mineshige}}]{2000PASJ...52..133W}
{Watarai}, K.-y., {Fukue}, J., {Takeuchi}, M., \& {Mineshige}, S. 2000, \pasj, 52, 133

\bibitem[{{Wijnands} {et~al.}(2012){Wijnands}, {Yang}, \& {Altamirano}}]{2012MNRAS.422L..91W}
{Wijnands}, R., {Yang}, Y.~J., \& {Altamirano}, D. 2012, \mnras, 422, L91

\bibitem[{Wilms {et~al.}(2000)Wilms, Allen, \& McCray}]{wilms2000}
Wilms, J., Allen, A., \& McCray, R. 2000, The Astrophysical Journal, 542, 914

\bibitem[{{Wu} {et~al.}(2010){Wu}, {Yu}, {Li}, {Maccarone}, \& {Li}}]{2010ApJ...718..620W}
{Wu}, Y.~X., {Yu}, W., {Li}, T.~P., {Maccarone}, T.~J., \& {Li}, X.~D. 2010, \apj, 718, 620

\bibitem[{{Xu} {et~al.}(2017){Xu}, {Garc{\'\i}a}, {F{\"u}rst}, {Harrison}, {Walton}, {Tomsick}, {Bachetti}, {King}, {Madsen}, {Miller}, \& {Grinberg}}]{2017ApJ...851..103X}
{Xu}, Y., {Garc{\'\i}a}, J.~A., {F{\"u}rst}, F., {et~al.} 2017, \apj, 851, 103

\bibitem[{{Zhang} {et~al.}(2021){Zhang}, {Altamirano}, {Uttley}, {Garc{\'\i}a}, {M{\'e}ndez}, {Homan}, {Steiner}, {Alabarta}, {Buisson}, {Remillard}, {Gendreau}, {Arzoumanian}, {Markwardt}, {Strohmayer}, {Neilsen}, \& {Basak}}]{2021MNRAS.505.3823Z}
{Zhang}, L., {Altamirano}, D., {Uttley}, P., {et~al.} 2021, \mnras, 505, 3823

\bibitem[{{Zoghbi} {et~al.}(2016){Zoghbi}, {Miller}, {King}, {Miller}, {Proga}, {Kallman}, {Fabian}, {Harrison}, {Kaastra}, {Raymond}, {Reynolds}, {Boggs}, {Christensen}, {Craig}, {Hailey}, {Stern}, \& {Zhang}}]{2016ApJ...833..165Z}
{Zoghbi}, A., {Miller}, J.~M., {King}, A.~L., {et~al.} 2016, \apj, 833, 165

\bibitem[{{{\.Z}ycki} {et~al.}(1999){{\.Z}ycki}, {Done}, \& {Smith}}]{1999MNRAS.309..561Z}
{{\.Z}ycki}, P.~T., {Done}, C., \& {Smith}, D.~A. 1999, \mnras, 309, 561

\end{thebibliography}
\end{document}